\documentclass[reprint, amsmath, amssymb, aps, prapplied, floatfix]{revtex4-2}

\usepackage{graphicx}
\usepackage{dcolumn}
\usepackage{bm}
\usepackage[hidelinks]{hyperref}
\usepackage{upgreek}
\usepackage{xfrac}
\RequirePackage{braket}
\usepackage{xcolor}
\usepackage[export]{adjustbox}
\usepackage{ragged2e}
\usepackage{ulem}
\usepackage{dirtytalk}
\usepackage{float}

\newcommand{\be}{\begin{equation}}
\newcommand{\ee}{\end{equation}}
\newcommand{\bea}{\begin{eqnarray}}
\newcommand{\eea}{\end{eqnarray}}

\usepackage{color}

\def\tcorr{T_\text{corr}}
\DeclareMathOperator{\sinc}{sinc}

\graphicspath{{figures/}}

\begin{document}

\preprint{APS/123-QED}

\title{Ramsey correlation spectroscopy with phase cycling using a single quantum sensor}

\author{Inbar Zohar}
\affiliation{Department of Chemical and Biological Physics, Weizmann Institute of Science, Rehovot 7610001, Israel}
\author{Santiago Oviedo-Casado}
\affiliation{Universidad Politécnica de Cartagena member of European University of Technology EUT+, Área de Física Aplicada, Departamento de Física Aplicada y Tenología Naval, 30202 Cartagena, Spain}
\affiliation{Escuela Superior de Ingeniería y Tecnología, Universidad Internacional de La Rioja, 26006 Logroño, La Rioja, Spain}
\author{Andrej Denisenko}
\author{Rainer St\"{o}hr}
\affiliation{3.\,Physikalisches Institut and ZAQuant, Universit\"{a}t Stuttgart, Stuttgart 70569, Germany}
\author{Amit Finkler}
\email{amit.finkler@weizmann.ac.il}
\affiliation{Department of Chemical and Biological Physics, Weizmann Institute of Science, Rehovot 7610001, Israel}
\date{\today}

\begin{abstract}
Magnetic spectroscopy at the nanoscale provides unique insights into material properties, structure, and dynamics. Quantum sensors, such as nitrogen-vacancy (NV) centers in diamond, are ideally suited for these length scales. However, they face significant limitations when detecting low-frequency signals due to their finite coherence times. Signals oscillating slower than the inverse coherence time of the sensor (1/$T_2^*$) are difficult to detect, as coherence decays before sufficient phase accumulation occurs.
Here we present RESOLUTE (Ramsey corrElation SpectroscOpy puLse seqUence wiTh phasE cycling), a spectroscopy protocol that overcomes coherence time limitations for low-frequency signal detection. RESOLUTE combines Ramsey measurements with correlation spectroscopy, storing accumulated phase as population imbalance during a correlation period ($T_\mathrm{corr} < T_1$) between two sensing periods. This approach produces a frequency filter that generates a new coherence time $T_2^p > T_2^*$, and shifts the frequency-matching condition from the coherence time to the correlation time, enabling detection of signals in the spectral region between $1/T_1$ and $1/T_2^p$, previously inaccessible.
We experimentally demonstrate that RESOLUTE extends the effective coherence time from $T_2^* = 0.38\,\upmu\mathrm{s}$ to $T_2^p = 5.1\,\upmu\mathrm{s}$, surpassing even Hahn Echo measurements, and the technique successfully detects $^{13}\mathrm{C}$ nuclear spin Larmor precession at magnetic fields as low as 49\,G, corresponding to frequencies of $\sim 50\,\text{kHz}$. We provide theoretical insight using Fisher information to characterize RESOLUTE's frequency estimation capabilities and compare its performance to existing protocols. Finally, we combine the extended coherence time offered by RESOLUTE with adiabatic pulses for robust spin control and phase cycling for effective DC signal extraction to obtain enhanced sensitivity to electron spins, demonstrating non-trivial improvements in detecting weak dipolar interactions essential for single-molecule imaging and quantum sensing applications.
\end{abstract}

\maketitle

\section{Introduction}

\begin{figure}[ht!]
    \centering
    \includegraphics[width=\columnwidth]{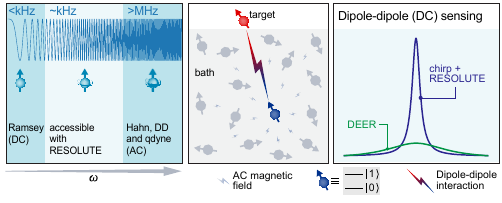}
    \caption{\textbf{Schematic of the experimental theater}. Left panel: The frequency domain of sensitivity to signals, with Ramsey covering very low frequencies limited by $1/T_2^*$, dynamical decoupling (DD) covering high frequencies and a gap which is covered by RESOLUTE, the pulse sequence we introduce below in the text (Sec.\,\ref{sec:Correlation}). Middle panel: A plot depicting the sensor qubit, the target qubit, having some dipole-dipole coupling between them, and a spin bath. Right panel: The combination of chirped pulses and our pulse sequence allows to significantly improve the readout contrast of target spins \textit{and} achieving a narrower linewidth (Sec.\,\ref{sec:Ad_Corr}).}
    \label{fig:schematic}
\end{figure}

Magnetic spectroscopy in nanoscale volumes aims at gauging the local environment in the proximity of the sensor. This is typically done in as large a dynamic spectral range as possible, as sketched in the left panel of Fig.\,\ref{fig:schematic}, thereby providing for a unique fingerprint of the material, which is instrumental in unlocking its properties, namely its structure, function and dynamics. Nanoscale approaches offer an alternative to traditional magnetic resonance ones, with interesting features such as single molecule sensitivity or the ability to operate at ambient conditions \cite{Cappellaro2015, Lukin2016, Cappellaro2017, Degen2017}. As a result, the array of techniques denoted as nanoscale nuclear magnetic resonance (NMR) are being adopted with increased fervor~\cite{Budakian2024}.

\begin{figure}[h!]
\includegraphics[width=\columnwidth]{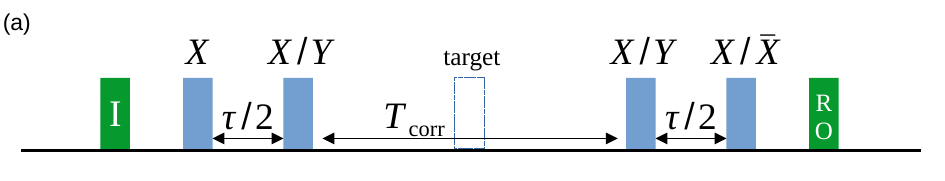}
\includegraphics[width=\columnwidth]{figures/fig2bc.jpg}
\caption{\label{fig:Background} \textbf{RESOLUTE pulses sequence, filter functions and sensitivity.} (a) Laser pulses for initialization and read-out of the NV center are in green, MW pulses for NV center driving in blue, and an MW pulse for the target electron spin in white with a dashed line. The different choices of letters give the full RESOLUTE with all phase combinations for signal extraction. (b) False-color map showing the sensitivity of the RESOLUTE filter function in Eq.\,\ref{eq:ff} as a function of the target signal frequency, $\omega$, and the correlation time, $T_\mathrm{corr}$. Here $\tau=5 \,\upmu\text{s}$. (c) A comparison between the sensitivity of Ramsey, Hahn Echo and RESOLUTE pulses sequences as a function of the frequency. The amplitude on each curve marks the sensitivity for fixed sequence parameters, while the color \textit{intensity} of each curve corresponds to the filter's ability to detect and isolate signals provided some sensor parameters. Thus, the stronger the color, the more sensitive the sequence is, with a decay in sensitivity (color strength) for lower frequencies resulting from the typical decay time of the sensor ($T_2^*$ = 0.36$\,\upmu\mathrm{s}$ for Ramsey and $T_2 = 4.3\,\upmu\mathrm{s}$ for a Hahn Echo). For RESOLUTE (the blue curve is the line-cut from panel a marked with a dashed line), the sensitivity window depends on the largest correlation time allowed $\tcorr < T_1$ (500$\,\upmu\mathrm{s}$ in the figure), which fixes the lowest frequency that can be detected, and the effective dephasing time $T_2^p$ (5.1 $\,\upmu\mathrm{s}$ in the figure) which in RESOLUTE represents a limit to the sensitivity to high frequencies. Note that all decay times considered are in agreement with our experimental measurements (see Fig.\,\ref{fig:T2_compare})}.
\label{fig:intro_filter_functions}
\end{figure}

Several table-top techniques exist nowadays, which allow one to map this wide magnetic spectrum. Notable among them are magnetic resonance force microscopy (MRFM), superconducting quantum interference devices (SQUID), electron spin resonance scanning tunneling microscopy (ESR-STM) and color centers in solids. For those that use two level (or resonance) systems as the sensing methodology--as displayed in the center panel of Fig.\,\ref{fig:schematic}--such as color centers (e.g., the nitrogen-vacancy center in diamond, or NV), it is their relaxation and decoherence properties that, more often than not, place a tighter bound on the available dynamic range \cite{SchaeferNolte2014}. For example, using a sensor with an intrinsic coherence time $T_2^*$ will preclude the detection of a signal that oscillates at a frequency lower than $1/T_2^*$, as by then the coherence will have decayed significantly (see Fig.\,\ref{fig:intro_filter_functions}b). Therefore, the low frequency regime gets defined through the intrinsic coherence time ($T_2^*$) of the sensor as the spectral region to which the sensor is less sensitive. 

Seeking to expand the dynamic range of resonance sensors, a host of control pulses sequences have been proposed and demonstrated over the years. These aim to improve the coherence time of the sensors, insulating them from unwanted noisy signals through frequency filtering, thereby expanding the accessible  frequency range (see for example the Hahn Echo range in Fig.\,\ref{fig:intro_filter_functions}b). Such sequences are gathered under the collective name of dynamical decoupling sequences \cite{Wang2012, Suter2016}, whose core principle is frequency matching \cite{Laraoui2011a}, whereby the pulses' time separation must be similar to the inverse frequency of the target signal. In these cases, low frequency signals are hard to detect, as the sensor loses coherence beyond recovery in between pulses. Additionally, aiming to increase the precision with which oscillating signals are detected, elaborate detection protocols such as correlation spectroscopy \cite{Laraoui2011a, laraoui2013high}, or phase sensitive measurements \cite{Schmitt2017, Degen2017, Glenn2018} are nowadays common. These allow to capture as much information as possible about the target signal within the designated experiment duration. Most of them, however, are unable to overcome the strict limit that the finite coherence time of the probe poses to the detection of signals oscillating at low frequencies with high enough sensitivity and sufficient frequency selectivity for efficient spectroscopy. Thus, access to a crucial region of the spectrum is severely restricted.

In this work, we present the Ramsey corrElation SpectroscOpy puLse seqUence wiTh phasE (RESOLUTE) sensing protocol for low frequency spectroscopy. Building upon ideas from Ramsey measurements and correlation spectroscopy, RESOLUTE is a relatively simple but powerful protocol that gives access to the low frequency regime with high sensitivity while filtering out noise sources. When also combining phase cycling, one can separate DC magnetic fields, to which Ramsey measurements are most sensitive, from AC magnetic fields correlated with the sequence correlation time, thereby offering higher sensitivity to the AC magnetic fields within the filter function of the sequence (see Fig.\,\ref{fig:intro_filter_functions}a). 

In RESOLUTE, the phase accumulated by the sensor during a first sensing period is stored as a population imbalance for a correlation time $\tcorr < T_1$, to be subsequently correlated with a second sensing period. By combining such measurement blocks, as shown in Fig.\,\ref{fig:intro_filter_functions}, RESOLUTE shifts the frequency matching condition to the correlation time, creating a frequency filter unrestricted by the coherence time of the probe. Thus, sensing coherent oscillations of nearby nuclear spins even at small bias magnetic fields ($<50\,\mathrm{G}$) becomes feasible. Moreover, the filtering feature of the correlation time means that the sensor is less affected by noisy signals, from which a new effective $T_2^p \gg T_2^*$ decoherence time emerges, that one can take advantage of to get a larger Ramsey sensing time. While some protocols exist to explore the very low frequency region of the magnetic spectrum \cite{Herbschleb2022, Oviedo2024}, they are based on the phase sensitive measurement scheme, which has a complicated laboratory implementation, a cumbersome post-processing of the data, and which does not provide any intrinsic frequency selectivity nor any change in $T_2^*$ unless dynamical decoupling is used. RESOLUTE, rather, offers a relatively simple sequence, a post-processing that requires no more than a least-squares fitting, and a natural frequency filter able to detect signals in noisy backgrounds with enhanced precision.

On top of the prolonged coherence, arising from the correlation, alternating the phases of the pulses in the main sequence building block shown in Fig.\,\ref{fig:intro_filter_functions}a following a phase cycling scheme (see Sec.\,\ref{sec:Correlation}) can offer further elimination of DC fields. Utilizing the phase cycling for DC extraction and the correlation time in the center as a window for flipping target DC interactions, specifically dipolar coupling interaction to other spins, one can actively force the DC dipolar coupling to be a correlated signal and focus the sequence on that field. Incorporating the manipulation of the DC interactions within the correlation window also allows for long manipulation protocols such as adiabatic pulses that are more robust and effective, as we demonstrate below. 

In what follows, we present a detailed experimental demonstration of the enhanced sensitivity that RESOLUTE offers to nearby nuclear spins, together with a theoretical analysis of the benefits and limitations that RESOLUTE has for spectroscopy. First, in Sec.\,\ref{sec:Correlation} we demonstrate that RESOLUTE overcomes the coherence time limitation by exploiting a correlation spectroscopy scheme that prolongs the $T_2^*$ coherence time of the sensor. We use the longer Ramsey measurements and sensitivity to oscillating signals, combined with phase cycling for DC field extraction, to detect carbon nuclear spins at low bias field via their Larmor precession. 
We then employ tools from quantum information theory in Sec.\,\ref{sec:Theory} to mathematically analyze the sequence and provide insights on the expected frequency estimation ability as a function of the sequence parameters, which allows us to pinpoint the spectral region in which RESOLUTE shines, and compare it with state-of-the-art protocols for low frequency detection. Finally, in Sec.\,\ref{sec:Ad_Corr} we use the longer sensing time enabled by RESOLUTE owing to the prolonged coherence time to demonstrate a non-trivial enhancement of the sensitivity to dipolar couplings. To do so, we combine RESOLUTE with phase cycling and adiabatic pulses, which, we show, are capable of overcoming the challenges inherent in a hard pulse due to the lower dependency that they have on the spin target orientation and the control that they allow over the pulse span without affecting its power. The enhanced sensitivity to dipolar couplings allows us to demonstrate detection of a target electron spin.

\label{new_introduction}

\section{\label{sec:experiments}Results}

\subsection{\label{sec:Correlation}Ramsey correlation sensing for prolonged coherence}

To address the limitation to frequency detection imposed by the coherence time of the sensor, we consider a correlation spectroscopy building block based on two Ramsey sensing periods of duration $\tau/2$ separated by a correlation period lasting for $T_\mathrm{corr}$. A full RESOLUTE measurement includes four repetitions of the sequence following a phase-cycled scheme, changing the phases of the MW pulses, as presented in Fig.\,\ref{fig:intro_filter_functions}a. The concept of a phase cycling sequence has been previously reported in electron spin resonance (ESR) \cite{kasumaj20085, kulik2002electron} and NMR \cite{Braunschweiler1983}. Recently, it has been experimentally implemented with NV centers \cite{Rovny2025}, but not in the context of electron or nuclear spin sensing.

Considering the ground state manifold of an NV center, a spin-1 system, assuming that a bias magnetic field lifts the degeneracy of the $\ket{\pm 1}$ states, and applying MW pulses that are resonant with the $\ket{0} \leftrightarrow \ket{-1}$ transition, the system Hamiltonian in the rotating wave approximation (RWA) with respect to the pulses reads 
\begin{equation}
    \mathcal{H}_{\mathrm{rot}} = \Delta B \sigma^{\mathrm{NV}}_z+H_S^{\mathrm{NV}},
    \label{eq:NV RWA}
\end{equation}
where $\Delta B$ includes any DC magnetic field or detuning from the rotating frame frequency, $\sigma_z^{\mathrm{NV}}$ is the Pauli matrix along the $\hat{z}$ quantization axis for the NV electron spin, and $H_S^{\mathrm{NV}}$ gathers all relevant environment interactions acting on the NV center probe (See Eq.\,\ref{eq:NV Hamiltonian} in Appendix \ref{ap:RCPS}).

Following initialization of the NV sensor in its $\ket{0}$ ground state through green laser illumination, the first Ramsey sensing period which initiates a RESOLUTE measurement begins with a $\left(\frac{\pi}{2}\right)$ rotation that creates a superposition state $\ket{\psi_\text{NV}}=\frac{1}{\sqrt{2}}\left(\ket{0}+\ket{1}\right)$, sensitive to  environmental interactions causing decoherence, and which evolves in time as $\ket{\psi_\text{NV}} = \frac{1}{\sqrt{2}}\left(\ket{0}+e^{i\Phi_1(t)}\ket{1}\right)$, accumulating a phase $\Phi_1(t)$ which is a function of any magnetic field interaction acting on the NV and oscillating slower than the interaction time. Thus, in our considered setting this phase gathers information about any dipolar coupling to the target spin ($\omega_{dd}$), MW detuning ($\Delta B$), hyperfine coupling ($A^N$), and slowly fluctuating noise ($\delta B$), such that after some Ramsey time $\tau/2$ the phase is $\Phi_{1}(\tau/2)=(\gamma_\mathrm{NV} \Delta B+\omega_{dd} + A^N + \gamma_\mathrm{NV} \delta B)\frac{\tau}{2}$, with $\gamma_\mathrm{NV}$ the gyromagnetic ratio for the NV electron spin. 

At the end of the first sensing period, the NV state is projected back along the $\hat{z}$ axis via a second $\left(\frac{\pi}{2}\right)_\phi$ pulse for the correlation time, in which the accumulated phase results in a population imbalance impervious to decoherence effects, where only a $T_1$ relaxation process limits the duration of this segment. Lastly, a second sensing period is initiated by another $\left(\frac{\pi}{2}\right)_\phi$ rotation, which creates a superposition state that depends on the phase accumulated during the first sensing period. The NV then accumulates a second phase, $\Phi_2(t)$, which again depends on the various decoherence interactions acting on the NV sensor. At the end of this second sensing period, a $\left(\frac{\pi}{2}\right)_\mathrm{RO}$ pulse projects the state back to populations and these are optically read via the NV fluorescence under green laser illumination, thus retrieving the information gathered by the sensor about its environment. 

The phase cycling in RESOLUTE corresponds to the use of different phases, either $\phi=x$ or $\phi = y$, for the two middle $\left(\frac{\pi}{2}\right)_{\phi}$ pulses, and alternating the sign $\left(\frac{\pi}{2}\right)_\mathrm{RO}$ with $\mathrm{RO} = \pm x$ for the final projection pulse for each of the possibilities in the middle pulses. Thus, a full RESOLUTE measurement is composed of four repetitions of the pulse sequence in Fig.\,\ref{fig:intro_filter_functions}a. While flipping the sign of the phase in the last pulse enhances contrast and suppresses systematic errors, alternating the phases on the middle pulses is key for isolating magnetic field components on resonance with the correlation time and eliminating DC magnetic fields influence. Depending on the phase $\phi$ of the middle pulses, the resulting spin state of the sensor after the final pulse is proportional to either the sine or the cosine functions of the total accumulated phase as
\begin{multline}
     S(\phi=x) = \braket{0|\psi_{\mathrm{NV}_{\phi=\hat{x}}}}=\bra{0} \cos\left(\Phi_1\right)\cos\left(\Phi_2\right)\ket{0}
     \\
     \shoveleft =\frac{\cos\left(\Phi_1-\Phi_2\right)+\cos\left(\Phi_1+\Phi_2\right)}{2}, 
     \\
     \shoveleft S(\phi=y) = \braket{1|\psi_{\mathrm{NV}_{\phi=\hat{y}}}} = \bra{1} \sin\left(\Phi_1\right)\sin\left(\Phi_2\right)\ket{1}
     \\ 
     \shoveleft=-\frac{\cos\left(\Phi_1-\Phi_2\right)-\cos\left(\Phi_1+\Phi_2\right)}{2}.\\
     \label{eq:projection_two}
\end{multline}
A detailed calculation of the sensor propagation during the RESOLUTE sequence can be found in Appendix\,\ref{ap:RCPS}.

For a static environment, the accumulated phases in both Ramsey periods are identical ($\Phi_1 = \Phi_2$). Then, when DC interactions remain unchanged between sensing periods, phase subtraction results in RESOLUTE being mostly insensitive to DC signals (See Fig.\,\ref{fig:RESOLUTE-XX} in Appendix \ref{ap:RCPS}). However, if magnetic fields change during the correlation time in between sensing periods, these changes persist upon subtraction. 
Consequently, adding the detected signal from the two cases of pulse phases ($\phi=x/y$, Eq.\,\ref{eq:projection_two}) permits isolating the magnetic fields that vary during the correlation period (Eq.\,\ref{eq:Ramsey_corr_tau}) in the accumulated phase on the NV. Thus, RESOLUTE filters out signals which are static with respect to $\tcorr$, thereby reducing the noise that affects the sensor. The fields that persist in the final signal can be classified in two groups: Correlated fields and fluctuating fields. The latter add noise, contributing to the effective coherence time $T_2^p$ of the sensor in RESOLUTE, while the former aid in stretching this coherence time beyond the natural Ramsey $T_2^*$ of the sensor. Crucially, these correlated fields can also themselves be a desired target signal, whose sensitivity is enabled by RESOLUTE.

\begin{equation}
\begin{split}
    S(\tau,T_\text{corr})_{\pm}&=S(\phi=x)\pm S(\phi=y)\\
    & =\frac{1}{2}\left[\cos(\Phi_1+\Phi_2)+\cos(\Phi_1-\Phi_2)\right]\\
    &\pm \frac{1}{2}\left[-\cos(\Phi_1+\Phi_2)+\cos(\Phi_1-\Phi_2)\right]\\
    &\propto \cos(\Phi_1\mp \Phi_2).
\end{split}
\label{eq:Ramsey_corr_tau}
\end{equation}

To better understand the effective NV center coherence time with RESOLUTE, let us consider first the $T_2^*$ decay time of Ramsey interferometry. In a Ramsey experiment, the $T_2^*$ decay time is a result of the averaging nature of the sequence with the NV center as a non single-shot-readout sensor \cite{Dobrovitski2008}. Then, if the NV center is subjected to a DC field, phase accumulation remains consistent across iterations, producing oscillations. However, if the field fluctuates faster than the measurement time, phase accumulation varies per iteration, contributing to signal averaging decay typically modeled by a Gaussian noise distribution (Eq.\,\ref{eq:Ramsey_decay}).  Longer interaction times that would allow sensing slower fluctuations, also amplify the variance of the noise, thus accelerating decay. Consider Ramsey detection of arbitrary fluctuating signals $\delta B$ as
\begin{equation}
    S=\int e^{-\alpha (\delta B)^2} \cos(\gamma_\mathrm{NV} \delta B t) d\delta B= e^{-\frac{\gamma_\mathrm{NV}^2 t^2}{4\alpha}}.
\label{eq:Ramsey_decay}
\end{equation}
Here, $\alpha$ reflects the inverse variance of the field fluctuations, and then $T_2^* = \frac{\sqrt{2\alpha}}{\gamma_\mathrm{NV}}$. 

In RESOLUTE, rather, the $\tcorr$ separation of two Ramsey periods suggests a natural splitting of the noise to two distinct regimes. The first consists of noisy signals that fluctuate slower than $\tcorr$, collectively denoted as $\delta B_{\mathrm{corr}}$. This noise contributes the same phase in both sensing periods, $\Phi_1\mathrm{(\delta B_{\text{corr}})}=\Phi_2\mathrm{(\delta B_{\text{corr}})}=\gamma_\mathrm{NV} \delta B_{\text{corr}} \frac{\tau}{2}$ and, therefore, gets canceled upon subtraction. Since this correlated noise is iteration-specific, it remains uncorrelated between $x/y$ phase shifts in a full RESOLUTE measurement, resulting in a noise-induced phase in the added phases term, resulting in 
\begin{equation}
\begin{split}
    S^{(+)}(\tau) = &\int e^{-\alpha_1 (\delta b+\delta B_\text{corr})^2} \left[S(\tau,T_\text{corr})_{+}\right] d(\delta b+\delta B_\text{corr})\\
    &\simeq e^{-\frac{\tau^2}{\alpha_1}}\cos(\Phi_1+ \Phi_2).
\label{eq:Ramsey_corr_DC_noise}
\end{split}
\end{equation}
The second regime includes all noise fluctuating faster than $T_\text{corr}$, denoted as $\delta b$, and which changes between the two sensing periods. Then, the phases are not canceled upon either phase subtraction of addition, resulting in 
\begin{equation}
\begin{split}
    S^{(-)}(\tau) =&\int e^{-\alpha_2 (\delta b)^2} \left[S(\tau,T_\text{corr})_{-}\right]d\delta b \\
    &\simeq e^{-\frac{\tau^2}{4\alpha_2}}\cos(\Phi_1- \Phi_2).
\label{eq:Ramsey_corr_noise}
\end{split}
\end{equation}
While $S^{(-)}(\tau)$ still decays due to uncorrelated noise between iterations, the decay is slower than a standard Ramsey signal (Eq.\,\ref{eq:Ramsey_decay}), as the correlation removes specific noise sources, which still affect the added signal.

\begin{figure}[t]
    \centering
    \includegraphics[width=1\linewidth]{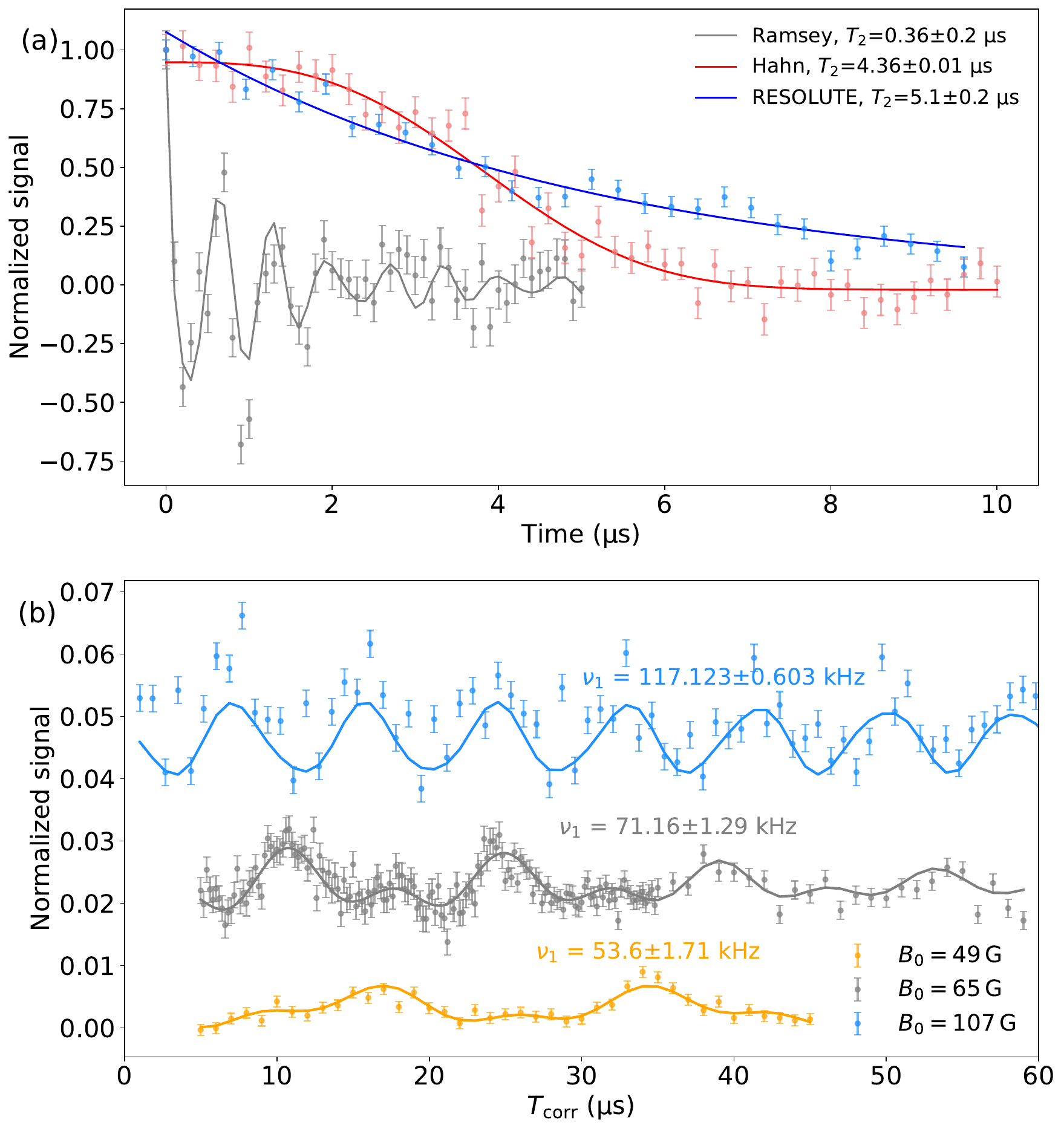}
    \caption{(a) Signal of RESOLUTE (blue), Hahn Echo (red), and Ramsey (gray) from a single NV showing the difference in decoherence times. The Ramsey signal also showed oscillations rising from the strong hyperfine coupling to the nitrogen nuclear spin. The RESOLUTE was taken with a correlation time of 10$\,\upmu$s (b) RESOLUTE signals with fixed sensing time $\tau=6\,\upmu\mathrm{s}$ and changing correlation time $\tcorr$, taken with $107.2 \pm 0.2$\,G (blue), $64.3 \pm 0.1$\,G (gray) and $49.5 \pm 0.3$\,G fields (orange), shifted vertically for clarity. All three show oscillations at the frequency of the carbon nuclear spin Larmor frequency.}
    \label{fig:T2_compare}
\end{figure}

With this theoretical understanding in mind, we measured the coherence time of a single NV center with a Ramsey sequence, a Hahn Echo sequence and RESOLUTE. The correlation time ($T_\mathrm{corr}$) used for RESOLUTE was $10\,\upmu\mathrm{s}$ and the bias magnetic field for all three scans was $64.3\pm 0.1\,\mathrm{G}$. The exponentially decaying signal measured in each sequence is plotted in Fig.\,\ref{fig:T2_compare}a. A short coherence time of $T_2^* = 0.38\pm 0.2\,\upmu\mathrm{s}$ was detected with the Ramsey spectroscopy (gray), on top of an oscillating signal rising from the strong hyperfine coupling to the nitrogen nuclear spin of the NV center \cite{Maze2012}. For RESOLUTE (blue line in Fig.\,\ref{fig:T2_compare}a), we obtain a coherence time $T_{2}^{p}=5.1\pm0.2\,\upmu\text{s}$, which is 15 times longer than the Ramsey $T_2^*$ owing to correlation of noise between the two sensing periods. Interestingly, the RESOLUTE coherence time $T_2^p$ is also longer than the $T_2$ extracted from a Hahn Echo measurement (red), indicating the presence of a noise source that is unaccounted for in the time scales of Hahn Echo, but \textit{is} refocused using RESOLUTE. 

Since the correlation time has a crucial role in filtering out noise from the signal, we have examined the RESOLUTE sequence by fixing the sensing time $\tau$ (see Fig.\,\ref{fig:intro_filter_functions}), while varying the correlation time $\tcorr$. In that manner, we are able to scan noise sources with different time scales or frequencies. For instance, an oscillating magnetic field with a low frequency $\omega_C$ such that its period time, $T_C$, is longer than a typical Ramsey or Hahn Echo sensing time ($T_C>\tau)$, would be a source of noise contributing an arbitrary phase to the signal and adding to the decay time (Eq.\,\ref{eq:Ramsey_decay}). However, in RESOLUTE, such a signal can be correlated if the correlation time is chosen properly to match the period time of the external field $T_{\text{corr}}=nT_{\mathrm{C}}$. 

In practice, the oscillating magnetic field contributes to the sensor phase in both sensing periods of RESOLUTE, such that $\Phi_1=\omega_C\frac{\tau}{2}$ and $\Phi_2=\omega_C\frac{\tau}{2}+\omega_CT_\text{corr}$. Therefore, when $T_\text{corr}=nT_C$ the phase shift rising during the correlation time is $2\pi$ and the phases of both periods are subtracted, eliminating the slow oscillating field noise from the signal. A detailed derivation can be found in Appendix \ref{ap:RCPS}.

Demonstrating this effect, we have measured RESOLUTE signals from the same NV with a constant sensing time of $\tau=6\,\upmu\mathrm{s}$ and varying correlation time $\tcorr$. We measured the signal under three different bias low magnetic fields, expected to induce a slowly-oscillating magnetic field arising from the Larmor frequency of the naturally abundant $^{13}\mathrm{C}$ nuclear bath. Figure \ref{fig:T2_compare}b presents the signals of all three fields, the gray curve for the field of $64.3\pm0.1\,\mathrm{G}$, the orange curve for the field of $B^{(0)}=49.5\pm0.2\,\mathrm{G}$, and the cyan curve for the field of $107.2\pm0.2\,\mathrm{G}$. 

Fitting the gray signal to a function of two cosines with a decay envelope reveals oscillations at frequencies of $\frac{\omega_1}{2\pi}=71.2\pm1.3\,\mathrm{kHz}$ and $\frac{\omega_2}{2\pi}=140.6\pm1.5\,\mathrm{kHz}$. The lower frequency, $\omega_1$, corresponds to the expected Larmor frequency of $^{13}\mathrm{C}$ nuclear spins in the lattice: $\frac{1}{2\pi}\omega_{^{13}\mathrm{C}}(B^{(0)}=64.3\pm 0.1\,\mathrm{G}) = 68.8\pm0.1\,\mathrm{kHz}$. The same reveals for the orange curve with revivals at a frequency of $\frac{1}{2\pi}\omega_1=53.2\pm1.6\,\mathrm{kHz}$, which matches the expected Larmor precession frequency of the carbon nuclear spin, $\frac{1}{2\pi}\omega_{^{13}\text{C}}(B^{(0)}=49.5\pm0.3\,\mathrm{G})=52.9\pm0.2\,\mathrm{kHz}$ and a second frequency of $\frac{1}{2\pi}\omega_2=116.8\pm2.9\,\mathrm{kHz}$, and for the cyan curve with oscillation frequency of $\frac{1}{2\pi}\omega_1 = 117\pm 1\,\text{kHz}$, corresponding to a Larmor precession frequency of $\frac{1}{2\pi}\omega_{^{13}\mathrm{C}}(B^{0}=107.2\pm0.2\,\mathrm{G})=114.5$\,kHz.

The second frequency in the fit with all three bias magnetic fields is approximately twice the $^{13}$C Larmor precession frequency. We attribute this to an inherent spurious effect of the correlation sequence (second harmonic) \cite{loretz2015spurious}, also indicated in the $\sinc$ nature of the filter function of the sequence (See Appendix\,\ref{AppendixTheory}). We note that the detection of the carbon nuclear spin Larmor precession frequency was reported before with a Hahn Echo correlation sensing sequence \cite{laraoui2013high}. However, that work stated that the signal cannot be detected with a bias magnetic field smaller than $110\,\mathrm{G}$, while in our case, this magnetic field range is accessible with RESOLUTE. 

\subsection{Theoretical analysis of RESOLUTE}\label{sec:Theory}

To gain understanding on the capabilities and limitations of RESOLUTE, we model the phase accumulated by a qubit interacting with a coherent external signal, and use it to calculate the Fisher information for RESOLUTE measurements. The Fisher information is an objective tool that allows to connect experimental results with mathematical modeling. Experimentally, measurements are carried out to get information about any given parameter of interest describing the target signal. The accuracy with which the parameter is measured can be quantified by the mean-squared-error of an estimation given the set of measurements. Theoretically, a physical model including the parameters of interest is used to describe the experiment. Mathematically, the smallest possible mean-squared-error for parameter estimation given the model is determined by the Crámer-Rao bound, and corresponds to the inverse Fisher information about said parameter in the proposed model \cite{Wootters1981, Braunstein1994}. Thus, the Fisher information can be used to estimate the expected performance of an experimental protocol and to fairly compare among different protocols.

As demonstrated in the previous section, RESOLUTE can be used to isolate correlated fields in the environment interacting with the probe. These can be modeled as pure tone signals oscillating at a frequency $\omega$, which we describe here by $A\sin(\omega t + \varphi)$. A classical figure of merit for sensitivity to external signals is the filter function of a given protocol, which corresponds to the signal's phase ($\varphi$) averaged square of the accumulated phase on the sensor. Therefore, we first need to calculate the accumulated phase on our NV center probe at the end of each of the four blocks in a RESOLUTE sequence. This accumulated phase corresponds to the integral of the signal during each of the two phase acquisition times of $\tau/2$ duration. As each block has a different combination of microwave pulse phases, the phase accumulated at the end of each block is a different combination of the phases accumulated during the first and second detection periods of a RESOLUTE block (see Fig.~\ref{fig:intro_filter_functions}). These differences become important in calculating the exact Fisher information, presented below. Yet at the same time, the full RESOLUTE sequence combination of four blocks with different pulses' phases can be well approximated by a global accumulated phase, providing intuition about signal detection with RESOLUTE. Combining the results from the four blocks, the effective accumulated phase is 
\be
\begin{split}
& \Phi(\tau,\tcorr,\omega) = \Phi_1(\tau,\tcorr,\omega) - \Phi_2(\tau,\tcorr,\omega) \\&= A\int_0^{\tau/2} \sin(\omega t + \varphi)dt  - A\int_{T_{\text{corr}}+\tau/2}^{T_{\text{corr}} + \tau} \sin(\omega t + \varphi)dt,
\end{split}
\ee
which, following some trigonometric identities, results in
\begin{multline}\label{accumphase}
\Phi(\tau,\tcorr,\omega) = A\tau\sin\left(\frac{\omega T_{\text{corr}}}{2} + \frac{\omega \tau}{2} + \varphi \right)\times \\ \cos\left(\frac{\omega T_{\text{corr}}}{2} + \frac{\omega \tau}{4}\right)\sinc\left(\frac{\omega \tau}{4}\right).
\end{multline}

We can use Eq.\,\ref{accumphase} to calculate the filter function for RESOLUTE, which corresponds to $\langle \Phi(\tau,\tcorr,\omega)^2 \rangle_\varphi$, the subscript $\varphi$ indicating averaging over the signal phase, which reflects real experimental conditions where each measurement witnesses a different, random value of $\varphi$. Then
\begin{equation}
\label{eq:ff}
\langle \Phi(\tau,\tcorr,\omega)^2 \rangle_\varphi = \frac{A^2\tau^2}{2}\cos^2\left(\frac{\omega T_{\text{corr}}}{2} + \frac{\omega \tau}{4}\right)\sinc^2\left(\frac{\omega \tau}{4}\right).
\end{equation}
Equation\,\ref{eq:ff} improves upon the Ramsey sequence filter function, which reads $\langle \Phi_R^2 \rangle \sim \sinc\left(\frac{\omega_{\tau_R}}{2}\right)$ and peaks at zero frequency, by adding a modulation that centers the spectral peaks around $\omega \tilde{T} = 2\pi n$ with $n \in \mathcal{N}$ and $\tilde{T} = T_{\text{corr}} + \tau/2$, showing that it is the interplay between $T_{\text{corr}}$ and $\tau$ that unlocks frequency selectivity for consecutive Ramsey sensing periods. In fact, Eq.\,\ref{eq:ff} closely resembles the filter function of a Hahn Echo (see Eq.\,\ref{eqSI:HEff} in Appendix~\ref{AppendixTheory}), in which the frequency matching condition that in a Hahn Echo is played by the pulses' separation time, is tackled in RESOLUTE by $\tilde{T}$. This, in turn, rationalizes the emergence of a new coherence time $T_2^p \gg T_2^*$, as RESOLUTE measurements act like a frequency filter that prevents the NV center from being sensitive to noisy signals well outside the range of the frequency matching condition. 
On Fig.\,\ref{figSI:theoryFF}, displayed in Appendix\,\ref{AppendixTheory}, we observe that Eq.\,\ref{eq:ff} reproduces well the position of the spectral peaks for three frequencies $\frac{1}{2\pi}\omega_1 = 117$\,kHz, $\frac{1}{2\pi}\omega_1 = 71$\,kHz and $\frac{1}{2\pi}\omega_2 = 53$\,kHz, using for the calculation the same parameters as for the experimental results shown in Fig.\,\ref{fig:T2_compare}b.

Focusing on the cosine factor of Eq.\,\ref{eq:ff}, we observe that it centers the peak around the target frequency, while the $\sinc$ factor narrows the corresponding spectral line. The peak amplitude is governed by $A\tau$ and therefore depends on the Ramsey sensing time, imposing a limitation on $\tau$, which can not be made arbitrarily small in order to match the desired target frequency. On the other hand, $T_{\text{corr}}$ needs to be greater than $\tau$ in order to have vanishing non-diagonal elements during the correlation time of the sequence. Both conditions imply that high frequency signals are hard to address with RESOLUTE, as they require small sensing time or balancing a small correlation time with the condition $T_{\text{corr}} > \tau$, which can be difficult to achieve while maintaining a reasonable phase accumulation. However, the same conditions mean that RESOLUTE is particularly useful to target low frequency signals, which require a large $\tilde{T}$, allowing for a greater sensing time $\tau$ and therefore more accumulated signal per measurement, while using $T_{\text{corr}}$ to match the resonant condition with the target signal's frequency, fulfilling at the same time the condition $T_{\text{corr}} > \tau$. 

Eq.\,\ref{eq:ff} also explains the difference between RESOLUTE and other Ramsey-based low frequency protocols, such as those proposed in Refs.\,\onlinecite{Herbschleb2022, Oviedo2024}, which follow the phase-sensitive approach \cite{Schmitt2017,Degen2017}. The reason stems from the behavior of the Ramsey filter function. A collection of Ramsey measurements loses information on the signal's phase, which leads to the filter function peaking at zero frequency, preventing detection of spectral lines at a finite frequency. To regain frequency sensitivity, one can perform Ramsey measurements which are evenly separated in time, thus coherently tracking the phase of the signal, preventing a uniformly distributed phase averaging, and allowing the correlation of measurements during post-processing, thereby unveiling the information about the frequency \cite{Herbschleb2022, Oviedo2024}. The alternative which RESOLUTE represents is to correlate measurements pairwise \textit{during} the measuring process by keeping the separation between them constant through $\tcorr$. Then, upon averaging of the phase, correlations survive which enable sensitivity to oscillating signals. The advantage that RESOLUTE offers is that by introducing the extra time-variable $\tcorr$  greater flexibility to match the target's signal frequency is gained, meaning that rather than relying on continuous measurement accumulation, RESOLUTE can be tailored to specific signals, thus becoming a more precise tool.

Moving to the Fisher information analysis of the protocol, we calculate it for our parameter of interest, which is the frequency of the target signal. We use the probability of a measurement outcome in the sensor, which reads
\be\label{singleFI}
\begin{split}
i_\omega &= \sum_{X\in\left\{0,1\right\}} \frac{1}{P(X|\omega)}\left[ \frac{dP(X|\omega)}{\omega}\right]^2 \\&= \frac{1}{P(0|\omega)[1-P(0|\omega)]}\left[ \frac{dP(0|\omega)}{\omega}\right]^2,
\end{split}
\ee
where we use the fact that for a binary measurement in a qubit probe $P(1|\omega) = 1-P(0|\omega)$. We can write the probability for the qubit to be in its ground state after a RESOLUTE block in terms of the accumulated phase, yielding  
\be
P(0|\omega) = \frac{1}{2}\left\{1+e^{(-t/T_d)}\sin\left[\Phi(\tau,\tcorr,\omega)\right]\right\},
\ee
where $T_d$ accounts for any environmental effect in the qubit. Then, each measurement block in a RESOLUTE sequence yields a Fisher information 
\be \label{phaseFI}
i_\omega = \frac{\sin^2\left[\Phi(\tau,\tcorr,\omega)\right]}{\left[e^{\left(2t/T_d \right)}-\cos^2\left[\Phi(\tau,\tcorr,\omega)\right]\right]}\left(\frac{d\Phi(\tau,\tcorr,\omega)}{d\omega}\right)^2.
\ee

Using the additivity of the Fisher information for independent trials \cite{Carlen1991} allows adding up all four measurement blocks to calculate the Fisher information in an RESOLUTE sequence and, further, the total Fisher information in a given experiment. As each block corresponds to a different projection on the Bloch sphere dependent on the phases at which the pulses are applied, the specific way the phases of the two accumulation periods combine changes from block to block, meaning that the exact expression for the Fisher information of the four blocks becomes quite involved, and using it for estimating the expected mean-squared-error of a specific experiment is difficult. In Appendix~\ref{AppendixTheory} we numerically explore the exact Fisher information for a RESOLUTE experiment with a wide range of parameters. In what follows, rather, we derive an approximate expression that reproduces well the exact behavior (see Fig.\,\ref{figSI:theory:exactVsApprox_tau5} in Appendix~\ref{AppendixTheory} for a comparison between the exact and the approximate Fisher information expressions) that can be used to easily understand the characteristics of RESOLUTE, and interpret its range of validity and applicability for frequency estimation.

To begin our derivation, note that despite the overall sign change, 
for the XXX($\pm$X) blocks the Fisher information per block is the same. That is also the case for the XYY($\pm$X) blocks. Furthermore, while the XXX($\pm$X) blocks depend on $\phi_2 - \phi_1$, the 
XYY($\pm$X) include a dependence on both $\phi_2 \pm \phi_1$. Whenever the accumulated phase in each sensing period is small, which is typically the case in experiments, the additive part $\phi_2 + \phi_1$ is quasi-static, and its derivative then approximates zero. Therefore, to a good approximation, we can simplify the Fisher information for a RESOLUTE sequence to four times the Fisher information of one of the blocks, e.g. the XXXX block. Then we have that
\be\label{eq:FI1}
i_\omega \approx \frac{4\sin^2\Phi}{\left[\exp\left(2\tau/T_2^p + 2T_{\text{corr}}/T_1 \right)-\cos^2\Phi\right]}\left(\frac{d\Phi}{d\omega}\right)^2,
\ee
where we consider that the qubit probe is subjected to decoherence with characteristic RESOLUTE time $T_2^p$ and relaxation with characteristic time $T_1$.

The RESOLUTE sequence requires $\omega \tau$ to be small, owing to the Ramsey fringe that the $\sinc$ term determines in Eq.\,\ref{accumphase}, which imposes $\sinc \approx 1$ for good signal accumulation. Then, we can approximate $\sinc' \approx 0$, for which reason the terms depending on the $\sinc$ derivative can be neglected. Following some trigonometry, we have that
\be\label{eq:phasederiv}
\left(\frac{d\Phi}{d\omega}\right)^2 \approx \frac{A^2\tau^2\tilde{T}^2}{4}\cos^2\left( \omega\tilde{T}+\frac{\omega \tau}{4}+\varphi\right)\sinc^2\left(\frac{\omega \tau}{4}\right),
\ee
with $\tilde{T} = T_{\text{corr}} + \tau/2$ as defined above. Additionally, the pre-factor 
\be
\frac{\sin^2\Phi}{\left[\exp\left(2\tau/T_2^p + 2T_{\text{corr}}/T_1 \right)-\cos^2\Phi\right]},
\ee in Eq.\,\ref{eq:FI1} oscillates with the accumulated phase $\Phi(\tau,\tcorr,\omega)$, and it is limited by the exponential factor, which imposes a limitation on the sensing  and correlation times. Considering the opposite oscillations of the sine and cosine terms, and the limits for $\tau$ and $T_{\text{corr}}$, the pre-factor can be bounded from above by $\exp\left(-2\tau/T_2^p - 2T_{\text{corr}}/T_1 \right)$, which reflects the limits on the different RESOLUTE times. Overall, it means that we can reasonably approximate the single measurement Fisher information by 
\be
\begin{split}\label{eq:approxFI}
i^{\mathrm{approx}}_\omega &= 8A^2\tau^2\tilde{T}^2\exp\left(-2\tau/T_2^p - T_{\text{corr}}/T_1 \right) \\& \times\cos^2\left( \omega\tilde{T}+\frac{\omega \tau}{4}+\varphi\right)\sinc^2\left(\frac{\omega \tau}{4}\right),
\end{split}
\ee
which agrees well with the exact calculated Fisher information for most frequencies of interest, as shown in Fig.\,\ref{figSI:theory:exactVsApprox_tau5}. Then, Eq.\,\ref{eq:approxFI} can be reliably used to determine the expected frequency estimation capabilities of an experiment, and to explore the possibilities that the RESOLUTE sequence offers. 

The first thing to notice in Eq.\,\ref{eq:approxFI} is that it fixes the frequency range accessible by the RESOLUTE sequence, which is insensitive to frequencies lower than $1/T_1$, due to the limitation in $\tcorr < T_1$. Sensitivity to large frequencies is limited by the condition $T_{\text{corr}} > \tau$ and the fact that neither $\tau$ nor $T_{\text{corr}}$ can be made arbitrarily small, as the $\tau \tilde{T}$ factor in Eq.\,\ref{eq:phasederiv} shows --- otherwise there is not enough contrast. Furthermore, phase accumulation is limited by the condition $\tau < T_2^p$ in the sensing time. Note that for typical $T_1$ values RESOLUTE allows to explore the lower part of the spectrum which is inaccessible to more involved sequences of the dynamical decoupling kind, which are limited by $\tau_{DD} < T_2 \ll T_1$ and therefore can only access frequencies that are larger than $1/T_2$. Moreover, the extended $T_2^* \rightarrow T_2^p$ decoherence time that RESOLUTE provides in comparison to other Ramsey sequences allows a larger phase acquisition time and means that, in the region in which RESOLUTE is sensitive, it provides a much larger sensitivity than other low frequency protocols based on the Ramsey sequence.  

Regarding the best experimental setup for a given target signal, Eq.\,\ref{eq:approxFI} tells us that low frequencies favor $\tau \sim T_2^p$, while larger frequencies require shorter phase acquisition times and, consequently, due to the $\tau^2$ factor in Eq.\,\ref{eq:approxFI}, are proportionally more difficult to detect, which is a direct consequence of the fact that we are using a Ramsey protocol, reflected in the $\sinc$ term on Eq.\,\ref{eq:approxFI}. Moreover, as a rule, the longer the correlation time, the better a lower frequency will be detected. The reason is that stretching the correlation time to its maximum before the signal loses coherence (decays) permits more oscillations of the target signal to imprint on the probe, which makes frequency estimation easier through, e.g., least-squares fitting. Note that here we are assuming that $\tcorr$ is just limited by $T_1$, when in a realistic experimental scenario, harder limits such as target signal inherent decay~\cite{Staudacher2015} or experimental constraints might exist which impose a tighter limit on $\tcorr$. Nevertheless, as the $\tilde{T}^2$ factor in Eq.\,\ref{eq:approxFI} demonstrates, the optimal strategy is to always use the largest $\tcorr$ possible. Finally, we observe that the optimal $\tau$ is approximately $T_2^p$, showing that the coherence time stretching that RESOLUTE provides is crucial for its ability to target small frequencies. As the exact calculation in Fig.\,\ref{figSI:theory:tauTcorrFIomega} on Appendix\,\ref{AppendixTheory} shows, a larger sensing time leads to the formation of two equally sensitive bands or branches in $\tcorr$, meaning there exists two different correlation times that yield similar estimation error, which is useful to note in $\tcorr$ limited scenarios. Moreover, accurately balancing the sensing time $\tau$ to meet the resonance condition for $\tilde{T}$ is key, and it might be advantageous to choose a shorter $\tau$, particularly at low frequencies. 

\begin{figure}
\includegraphics[width=\columnwidth]{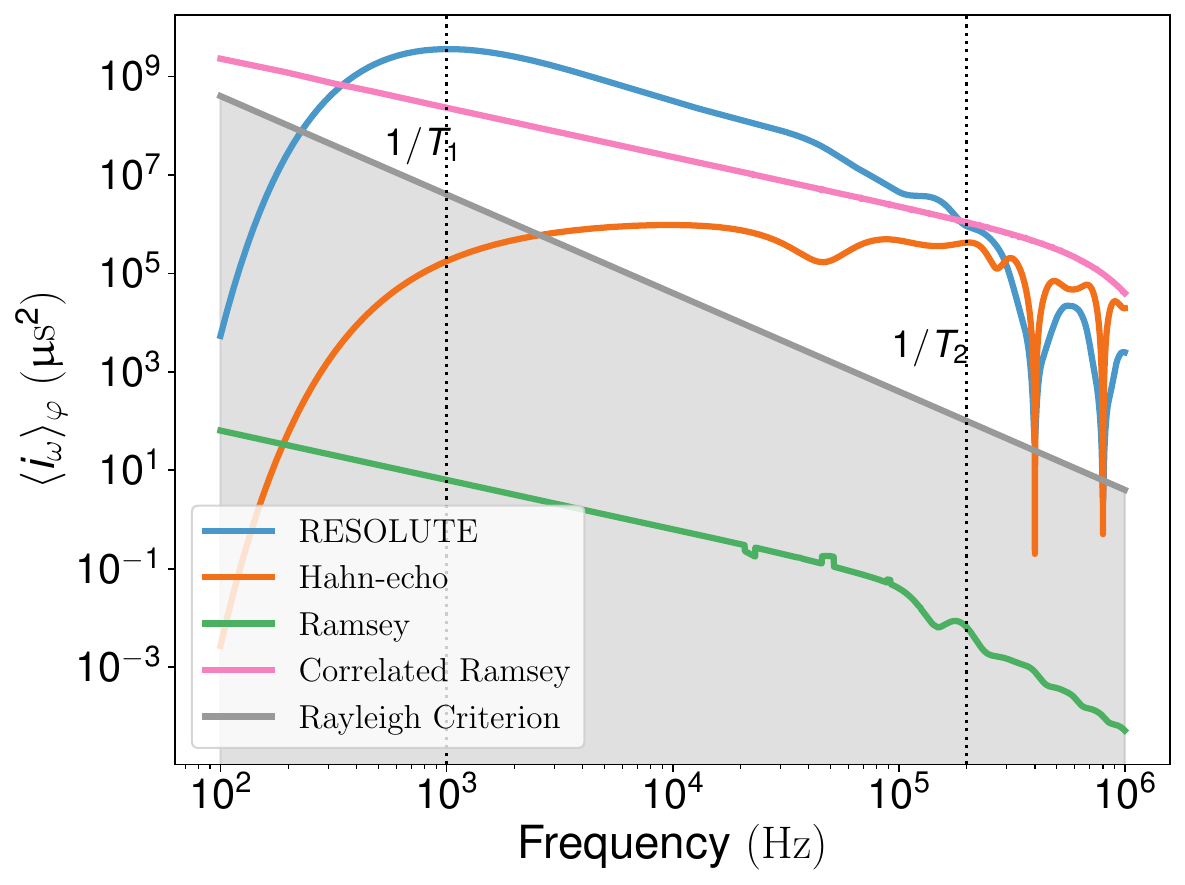}
\caption{Accumulated Fisher information for frequency detection for 500 repetitions of a RESOLUTE sequence, as a function of the target signal frequency. For comparison, we include the equivalent Fisher information using a Hahn Echo sequence, a simple Ramsey sequence, and the Correlated Ramsey measurements proposed in Refs.\,\onlinecite{Herbschleb2022, Oviedo2024}. For both RESOLUTE and the Hahn Echo, we assume that the inverse frequency is matched by either $\tcorr$ for RESOLUTE or $\tau_H$ in the case of the Hahn Echo. In both cases, we consider a $T_2^p$ ($T_2$ for the Hahn Echo) of $5\,\upmu\text{s}$. For the Ramsey and Correlated Ramsey, we consider a $T_2^* = 0.5 \,\upmu\text{s}$ and an optimal phase accumulation time $\tau_R = T_2^*$. In all cases, $T_1 = 1000 \,\upmu\text{s}$, and the overhead time in between measurements is $3\,\upmu\text{s}$. The shadowed area marks the region in which information is not sufficient for successful frequency estimation, delimited by the Rayleigh criterion from optics at $\Delta \omega^2 > 4/\omega^2$.}\label{fig:theory:totalFI}
\end{figure}

To complete our analysis, we calculate the Fisher information for frequency estimation in a given experiment of total duration $T_{\text{tot}} = 4 N\tilde{T}$ composed of $N$ repetitions of a RESOLUTE sequence. Since each sequence is uncorrelated with the rest, the phase $\varphi$ is randomly sampled each time, and consequently, the information accumulation is a direct sum for each block on the sequence, to whose result we must perform a phase averaging. Using then Eq.\,\ref{eq:approxFI} we can approximate the Fisher information for frequency detection for a given experiment by 
\be
i_\omega \approx 4A^2\tau^2\tilde{T} T_{\text{tot}}4e^{-2\tau/T_2^p}e^{- 2T_{\text{corr}}/T_1},
\ee
showing that RESOLUTE provides linear accumulation of information in time for frequency estimation. 

To demonstrate the frequency detection capability of RESOLUTE and compare it to alternative protocols for low frequency detection, in Fig.\,\ref{fig:theory:totalFI} we calculate the exact Fisher information for an experiment comprising 500 RESOLUTE sequences, and compare it for an equivalent experiment utilizing a Hahn Echo, a simple Ramsey sequence, or the Correlated Ramsey protocol from Refs.\,\onlinecite{Herbschleb2022, Oviedo2024}. For an equal experiment duration, RESOLUTE provides better frequency estimation ability within the range of applicability, namely for $1/T_2^p > \frac{1}{2\pi}\omega > 1/T_1$, as quantified by the Fisher information. We also include the frequency estimation limit defined by the Rayleigh criterion from optics \cite{Rayleigh1879} to be $\Delta \omega = 4/\omega^2$, showing that simple Ramsey accumulation is not suitable for frequency detection, while a Hahn Echo is only valid for large frequencies, being limited by the sensing time $\tau_{\text{H-E}} < T_2$. Overall, this analysis demonstrates the optimal range of applicability of RESOLUTE, in which it proves a superior protocol for frequency detection, and supplies a simple formula with which to estimate the best experimental parameters for targeting a given oscillating signal of interest.

\subsection{\label{sec:Ad_Corr}Overcoming the limits of single-electron spin sensing}

The longer coherence time that RESOLUTE unlocks permits larger Ramsey sensing periods, which means RESOLUTE has larger sensitivity compared to conventional Ramsey and some dipolar decoupling sequences, opening a wide range of new possibilities such as sensing of static weak signals from nuclear and electron spins. For example, one potential application is detecting the dipolar coupling of electron spins with the sensor. RESOLUTE is, by design, insensitive to DC signals, therefore, we need to recover DC sensitivity by acting on the target signal. To do so for electron spins, we apply a driving field to the target spin at its Larmor precession frequency~\cite{Grotz2011}, as employed in the double electron-electron resonance (DEER) pulse sequence (see Appendix \ref{ap:DEER}). To enhance detection, the field has to be optimized for driving the electron spin in a $\pi$-pulse ($\sigma_x$ gate) or utilize an adiabatic driving pulse, i.e., a chirped pulse. These pulses are broader in frequency and can overcome the uncertainty in the parameters of the $\pi$-pulse (experiment and theory supporting this claim are discussed in Appendix \ref{AppendixE}). To achieve an optimal chirp pulse, the adiabaticity of the pulse has to be optimized by considering the pulse duration, frequency span and the Rabi frequency of the spin to calculate the adiabaticity factor $Q$ that should be larger than 1 (see Eq.\,\ref{eq:Q min} in Appendix \ref{AppendixE}).

\begin{figure*}
\includegraphics[width=0.95\textwidth]{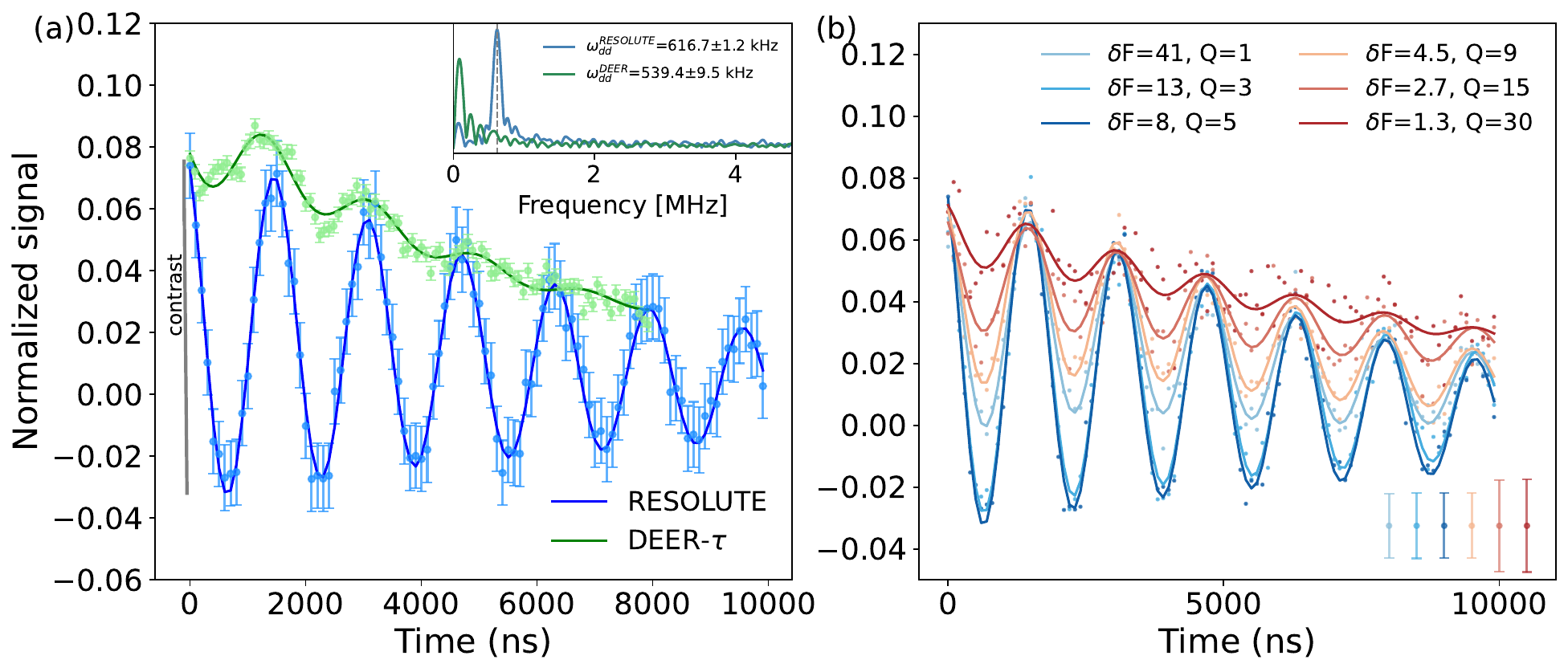}
\caption{\textbf{The effect of chirped pulses}. (a) RESOLUTE signal (blue) and DEER time signal (green), both taken with $2\,\mathrm{\upmu s}$ long chirp pulse with $Q=5$, as detected from an NV center with a strong dipolar coupling to an electron spin. The inset shows the fast Fourier transform of the both signals and the detected dipolar coupling as extracted from the fitted data. (b) RESOLUTE signal with fixed duration of $2\,\mathrm{\upmu s}$ chirp pulse and different adiabaticities as detected from the same NV-electron system reported in panel a. Error bars of each signal are presented in the lower right side of the plot.}
\label{fig:Rams_corr_chirp}
\end{figure*}

To combine the chirp pulse with the RESOLUTE sequence for dipolar coupling sensing, we apply the pulse during the correlation time (Fig.\,\ref{fig:intro_filter_functions}) while the correlation time is fixed to the relevant period time of the carbon nuclear spin bath ($T_C$) for noise cancellation and sensor coherence enhancement. The goal of the electron spin pulse is to flip the state of the target spin between the two sensing periods, effectively causing the DC magnetic field of the target dipolar coupling to flip. Thus, all other DC phase terms cancel upon subtraction (Eq.\,\ref{eq:Ramsey_corr_tau}), leaving only the phase arising from the dipolar coupling, expressed as $S^{(-)}(\tau,T_\text{corr})=A \cos(\omega_{dd}\tau)$ (full derivation is given in Appendix \ref{ap:RCPS}). The extended correlation period allows for a sufficiently wide chirped pulse, ensuring an optimal driving of the target spin (see Appendix \ref{AppendixE}). 

To compare the performance of RESOLUTE with a standard DEER-based pulse sequence for sensing dipolar couplings (denoted here as DEER-$\tau$, see Refs.\,\onlinecite{Shi2015, Schlipf2017} and Appendix \ref{ap:DEER_t}), we applied a chirped pulse using a $2\,\mathrm{\upmu s}$ pulse with adiabaticity of $Q=5$ (see Appendix \ref{AppendixE} for the precise definition of $Q$) instead of a $\pi$-pulse. The RESOLUTE signal resulted in oscillations with a high contrast of $C=6\,$\% of the normalized signal (see Fig.\,\ref{fig:Rams_corr_chirp}a for a visual definition of the contrast), and also leads to a low error in the estimated dipolar coupling extracted from the fitted signal and a strong Lorentzian signal around the dipolar coupling in frequency domain given by a Fourier transform (see inset in Fig.\,\ref{fig:Rams_corr_chirp}a). These results are starkly different from the DEER-$\tau$ sequence, which produces oscillations at $\frac{1}{2\pi}\omega_{dd}=0.6\,\mathrm{MHz}$ with a significantly lower contrast of $C=1.4\,$\% of the normalized signal and a higher fit error and a signal mostly masked by spurious noisy peaks in the frequency domain.

Given these findings, we further investigated the effect of chirp parameters on the RESOLUTE signal. Figure \ref{fig:Rams_corr_chirp}b presents RESOLUTE signals for a fixed pulse duration ($T_p=2\,\mathrm{\upmu s}$) with varying adiabaticity factors~($Q$). The highest dipolar coupling oscillation contrast, $C=5.9\pm0.1\,$\%, was achieved with $Q=5$. As $Q$ decreases (green and blue curves), the contrast decreases as well, with $C=3.2\pm0.1\,$\% for $Q=1$. This reduction was expected, as lower adiabaticity reduces pulse effectiveness.

Intuitively, increasing the adiabaticity factor should enhance contrast. However, as shown in Fig.\,\ref{fig:Rams_corr_chirp}b, the contrast decreased at higher values (pink, orange, and red curves). This effect arises because higher values reduce the total pulse span, thereby decreasing its effective range and spin-flipping efficiency.

\section{\label{sec:Discussion}Discussion}
Despite the relatively wide range of spectroscopic tools available for nanoscale magnetic sensing, the limit of low frequencies $\frac{1}{2\pi}\omega  < 1/T_2^*$ has generally suffered from poor sensitivity. Our pulse sequence, RESOLUTE or phase-cycled Ramsey correlation spectroscopy, is specifically tailored in such a way that atomic-sized sensors like the NV center in diamond can detect magnetic fields oscillating in the low frequency range. Thus, using RESOLUTE, we show detection of the magnetic field generated by a surrounding nuclear spin bath of $^{13}$C at fields as low as 49\,G. Moreover, we emphasize that this is not the lowest bound but, provided one can compensate for stray magnetic fields such as Earth's and other sources of low-frequency noise, a Fisher information based theoretical analysis demonstrates that RESOLUTE can be used to measure low-frequency oscillations in fields which are as low as 0.5\,G, limited only by the NV's $T_1$ time. Since the latter is a phonon-based relaxation mechanism, we can reasonably expect RESOLUTE to excel even more at low-temperatures, where phonons' contribution is less prominent \cite{Abobeih2018}. Moreover, RESOLUTE's $\omega > 1/T_1$ limitation to low frequency detection could be sidestepped by resorting to an external memory qubit with relaxation time $T_M > T_1$ to which the phase is mapped during the correlation time \cite{Zaiser2016}, thereby gaining access to even lower frequencies.

Furthermore, our theoretical analysis demonstrates that within the range of applicability, namely targeting frequencies in the range $1/T_2^p > \frac{1}{2\pi}\omega > 1/T_1$, RESOLUTE is likely the best existing candidate protocol, with the additional advantage of a simplicity of implementation and data analysis, and the added benefit of an enlarged coherence time of the sensor, which despite using Ramsey phase accumulation periods, lasts for $T_2^p \gg T_2^*$. Our analysis rationalizes this behavior as RESOLUTE acting like a frequency filter which makes the sensor impervious to specific noise signals. Then, RESOLUTE can be used as well as a very efficient DC field detector, owing to a longer Ramsey sensing time. Combined with chirped pulses, we show that RESOLUTE exhibits a dramatic improvement in the detection of electron spins, where the conventional protocols sometimes draw a blank. Finally, RESOLUTE could then be used to resolve the unique hyperfine interactions from each different $^{13}\mathrm{C}$ spins coupled to the NV center~\cite{Abobeih2019}. 

\begin{acknowledgments}
We are grateful for fruitful discussions with Nino Wili, Vadim Vorobyov and Durga Dasari, and Eyal Laster for proofreading the manuscript. This research is made possible in part by the historic generosity of the Harold Perlman Family. AF is the incumbent of the Elaine Blond Career Development Chair and acknowledges financial support from the Israel Science Foundation (Grants 418/20 and 419/20). S.O.-C acknowledges the financial support from the Agencia Estatal de Investigación (QNAVIUM Project SCPP2400C011413XV0 funded by MICIU/AEI/10.13039/501100011033).
\end{acknowledgments}

\appendix

\section{Methods}
We used three different diamond samples. For all RESOLUTE measurements of $^{13}\mathrm{C}$ Larmor precession measurements, we used a sample denoted here as `JWR2'. For the measurements on the electron spin using chirped pulses combined with RESOLUTE, we used two samples, denoted here as `205-4' and `AM-E2'. All three samples were cut from \textit{the same} original stock diamond, then thinned down to approximately $50\,\upmu\mathrm{m}$ thickness and polished to a surface roughness of $\mathrm{Ra}=1\,\mathrm{nm}$. Prior to etching, we used Ar/SF$_6$ for strain relief \cite{Momenzadeh2016} with an O$_2$ ``soft'' ICP finish \cite{Oliveira2015}. Subsequently, we etched nanopillar patterns into the diamond to increase photon collection efficiency~\cite{Momenzadeh2015}. Below is a table with detailed NV creation parameters:
\begin{table}[h!]
    \centering
    \begin{tabular}{c|c|c}
         \hline
         Sample & Energy & Dose \\
         \hline
                &          &                                     \\
         JWR2   & 5 keV    & $6\cdot 10^9\,\mathrm{cm}^{-2}$     \\
         205-4  & 9.8 keV  & $8\cdot 10^9\,\mathrm{cm}^{-2}$     \\
         AM-E2  & 9.8 keV  & $8\cdot 10^9\,\mathrm{cm}^{-2}$     \\ 
         \hline
    \end{tabular}
    \caption{Implantation parameters for the three samples used in this work. Typically a 5\,keV energy results in an implantation depth of $8\pm3$\,nm while the 9.8\,keV with $14\pm 5$\,nm. In all three samples, the dose was optimized to yield a final NV concentration of 1\,$\upmu\mathrm{m}^{-2}$.}
    \label{tab:placeholder}
\end{table}
All measurements were performed on a custom-built confocal fluorescence microscope \cite{Zohar_2023}. Pulses were orchestrated using the QM-OPX and the experiment was run using QuDi \cite{Binder2017}.

\subsection{DEER-frequency}\label{ap:DEER}
\begin{figure}[h!]
        \includegraphics[width = \columnwidth]{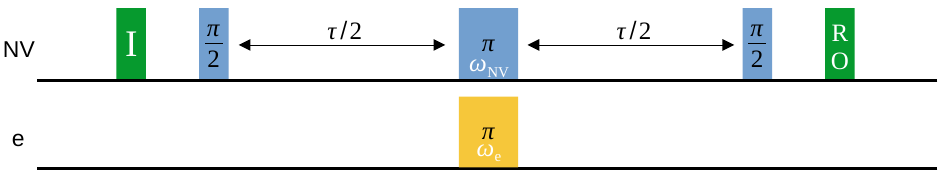}
        \caption{DEER-frequency pulse sequence. For a fixed spin-echo time, $\tau$, the frequency of a $\pi$-pulse on the target electron spin, $\omega_\mathrm{e}$, is scanned in order to find the resonant frequency corresponding to its Larmor precession frequency, $\gamma_e B_0$.}
        \label{fig:deer-frequency}
    \end{figure}

DEER-frequency, or DEER for brevity, is a double-electron-electron-resonance pulse sequence used to identify the resonance frequency of a target electron spin, $\omega_\mathrm{e}$, by nesting a $\pi$-pulse on this spin in a standard spin-echo pulse sequence on an NV center with a fixed $\tau$ between pulses, whose resonant frequency is $D-\gamma B_0 = \omega_\mathrm{NV}$ for the $m_s=-1$ spin substate (see Fig.\,\ref{fig:deer-frequency}).
\subsection{DEER-$\tau$}\label{ap:DEER_t}
Once the target electron's resonant frequency is identified, one can also fix $\omega_\mathrm{e}$ and scan the interaction time $\tau$. This modulates the coupling between the NV and the target electron spin, resulting in coherent oscillations at a frequency proportional to the coupling strength. In the main text, this is called a DEER-$\tau$ measurement.
\section{NV center sensing of slow AC fields and DC fields with RESOLUTE}
\label{ap:RCPS}

The NV center's interaction with the environment can be summarized in the rotating wave approximation Hamiltonian given in Equation \ref{eq:NV Hamiltonian}. The interactions are separated into two regimes: DC magnetic fields, including slow changing magnetic noise $\delta$B, dipolar coupling to electron spins $\omega_{dd}$ with $\sigma_z^e$ the Pauli operator along the $\hat{z}$ quantization axis of the coupled electron spin, and hyperfine coupling to the nuclear spin of the $^{15}\mathrm{N}$ atom, and AC magnetic fields, including Larmor precession of any spin in the vicinity of the probe NV center, such as $^{13}\mathrm{C}$ and $^{15}\mathrm{N}$ nuclear spins and electron spins.

\begin{equation}
\begin{split}
    \mathcal{H}_S&=\underbrace{\gamma_\text{NV}\delta B \sigma_z^\mathrm{NV}+\omega_{dd} \sigma_z^\mathrm{NV}\sigma_z^e+A_{zz}^{^{15}\text{N}}\sigma_z^\mathrm{NV}I_z^{^{15}\text{N}}}_{\text{DC\,\,fields}}\\
    &+\underbrace{\gamma_\mathrm{e}B_z\sigma_z^e+\gamma_\mathrm{^{13}\text{C}}B_zI_z^{^{13}\text{C}}+\gamma_\mathrm{^{15}\text{N}}B_zI_z^{^{15}\text{N}}}_{\text{AC\,\,fields}}
\end{split}
\label{eq:NV Hamiltonian}
\end{equation} 

The RESOLUTE pulse sequence is a method offering sensitivity to both DC magnetic fields and low-frequency AC fields upon different application of the sequence.

\subsection{DC sensing with RESOLUTE and dipolar coupling isolation}

The RESOLUTE pulse sequence consists of four $\frac{\pi}{2}$ pulses applied on the NV center and repeated in a phase-cycled manner \cite{Haeberle.2015, Rovny2025}. To understand the effect of DC fields on the final outcome of the sequence, one can describe the state of the sensor at the RWA after every pulse. 

After the first pulse, the NV center is at a superposition state $\ket{\psi_\text{NV}}=\frac{1}{\sqrt{2}}\left(\ket{0}+\ket{1}\right)$. During the first free evolution time $\sfrac{\tau}{2}$, the state of the NV center accumulates a phase according to the Hamiltonian of the system, and the NV center state gains a phase- $\ket{\psi_\text{NV}} = \frac{1}{\sqrt{2}}\left(\ket{0}+e^{i\phi_1}\ket{1}\right)$, where $\phi_1=\frac{\tau}{2} \left(\gamma\Delta B + A_n + \gamma \delta B+\omega_{dd}\left|\sigma_z^e\right|\right)$. The operator $\left|\sigma_z^e\right|$ will result in one of the eigenstates of the target electron spin- $\pm\frac{1}{2}$ reflecting the state of the electron during this evolution time. We take the absolute value of the eigenstate because of the symmetry of the measurement, where we will show that the interaction flips a sign in the second part of the pulse sequence.

The second pulse projects the state to the read-out axis. The state just before projection has the accumulated phase during the interaction time (Eq.\,\ref{eq:interaction_one}). After the second pulse, the spin is projected to one of the eigenstates of the sensor, and the phase translates to a change in the amplitude of the state (Eq.\,\ref{eq:projection_one}). The phase $\theta$ is that of the second pulse (for phase $\hat{x}\rightarrow\theta=0$ and for phase $\hat{y}\rightarrow\theta=\pi/2$, see Fig.\,\ref{fig:intro_filter_functions}). The remaining part of the spin  that was not projected (left on the equator) will keep accumulating phase during the correlation time. However, as the mixing time is longer than the decoherence time ($T_2^p$), it will disperse. The relaxation time $T_1$ of the sensor is typically two orders of magnitude longer than the mixing time. Therefore, we can neglect its effect in the ensuing analysis. Considering all these effects, the state during the mixing time will be as written in Eq.\,\ref{eq:projection_one}.

\begin{subequations}
    \begin{equation}\begin{split}
    \ket{\psi_\text{NV}}=
     & \cos\left(\theta\right)\left[\begin{split}&e^{T_{\text{corr}}/T_2^p}\sin\left(\phi_1\right) \left(\ket{0}+e^{i\phi_{\text{corr}}}\ket{1}\right)\\ &+e^{T_{\text{corr}}/T_1}\cos\left(\phi_1\right) \ket{1} \end{split}\right]\\
     & + \sin\left(\theta\right)\left[\begin{split}&e^{T_{\text{corr}}/T_2^p}\cos\left(\phi_1\right) \left(\ket{1}\right.\\ &+\left.e^{i\phi_{\text{corr}}}\ket{0}\right)+e^{T_{\text{corr}}/T_1}\sin\left(\phi_1\right) \ket{0}\end{split}\right]\end{split}
    \label{eq:interaction_one}
    \end{equation}
    \begin{equation}
    \ket{\psi_\text{NV}}=\cos\left(\theta\right)\cos\left(\phi_1\right) \ket{1} + \sin\left(\theta\right)\sin\left(\phi_1\right) \ket{0}
    \label{eq:projection_one}
    \end{equation}
\end{subequations}

The third pulse drives the sensor again to superposition but with a smaller amplitude and different phase, 
$$\ket{\psi_\text{NV}}=\left[\cos\left(\phi_1\right)\cos\left(\theta\right)+\sin\left(\phi_1\right)\sin\left(\theta\right)\right]\left(\ket{0}-e^{-i\theta}\ket{1}\right)$$
In this state, the sensor again accumulates a phase,
\begin{multline}
    \ket{\psi_\text{NV}}=\left[\cos\left(\phi_1\right)\cos\left(\theta\right)+\sin\left(\phi_1\right)\sin\left(\theta\right)\right]     \\
    \shoveleft\times\left(\ket{0}-e^{i\phi_2}e^{-i\theta}\ket{1}\right),\\ \nonumber
\end{multline}
where the second phase is $\phi_2=\frac{\tau}{2} \left(\gamma\Delta B+ A_n + \gamma \delta B \pm\omega_{dd}\left|\sigma^e_z\right|\right)$. While the sign of the dipolar interaction depends on the state of the target electron spin during the second sensing time. Finally, the fourth pulse is applied to project the state back to the read-out axis using a $\pi/2$ pulse with an $\hat{x}$ phase.

The outcome of the measurement depends on the phases of the second and third pulses. The state for each of the cases before projecting it to the read-out axis is presented in Eq.\,\ref{eq:Interaction_two}.
\begin{subequations}
    \begin{equation}
    \begin{array}{cc}
     &  \ket{\psi_{\mathrm{NV}_{\hat{x}\hat{x}}}}=\cos\left(\phi_1\right)\left(\ket{0}-e^{i\phi_2}\ket{1}\right)\\
     & \\ &\ket{\psi_{\mathrm{NV}_{\hat{y}\hat{y}}}}=\sin\left(\phi_1\right)\left(\ket{0}+ie^{i\phi_2}\ket{1}\right)
    \end{array}
    \label{eq:Interaction_two}
    \end{equation}
    \begin{equation}
    \begin{array}{cc}
     & \ket{\psi_{\mathrm{NV}_{\hat{x}\hat{x}}}}=\cos\left(\phi_1\right)\cos\left(\phi_2\right)\ket{0}\\
     & \\
     & \ket{\psi_{\mathrm{NV}_{\hat{y}\hat{y}}}}=\sin\left(\phi_1\right)\sin\left(\phi_2\right)\ket{1}
    \end{array}
    \label{eq:Projection_two}
    \end{equation}
\end{subequations}

\begin{figure}[t]
    \includegraphics[width = \columnwidth]{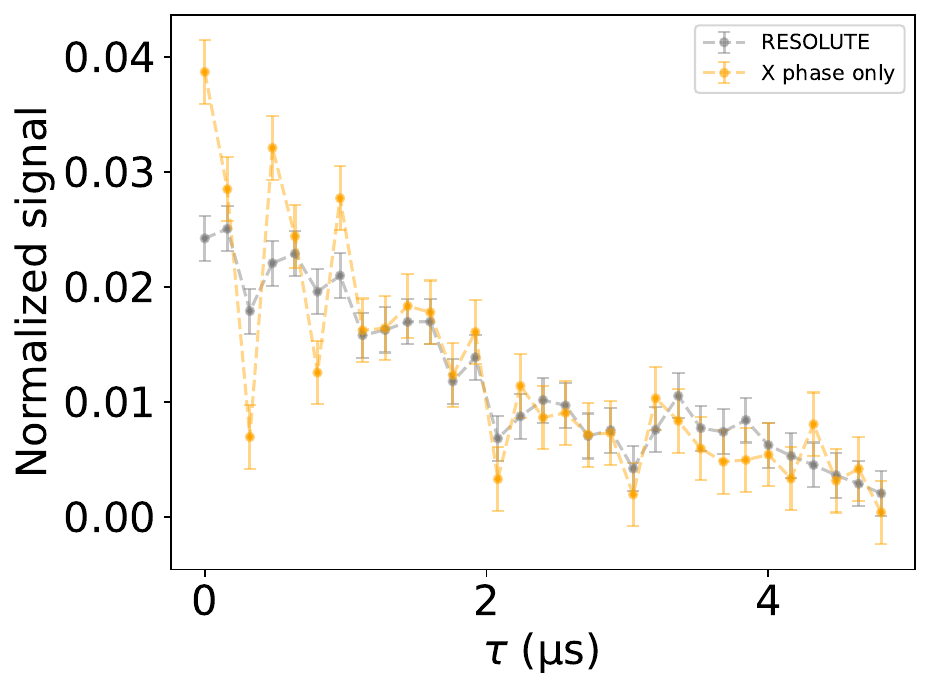}
    \caption{A full RESOLUTE pulse sequence with all four phase cycling cases (Eq.\,\ref{eq:Ramsey_corr_DC_noise}, gray) and only X phase signal (Eq.\,\ref{eq:projection_two}, yellow). Data taken with a fixed correlation time $T_{\text{corr}}$.}
    \label{fig:RESOLUTE-XX}
\end{figure}

Projecting these states to the read-out axis will result in one of the eigenstates of the NV center (Eq.\,\ref{eq:Projection_two}). After read-out, the signal would have a cosine or sine of the accumulated phase. Subtracting and adding the two outcomes and using a trigonometric identity will give two outcomes in Eq.\,\ref{eq:Ramsey_corr_tau}. Figure \ref{fig:RESOLUTE-XX} shows the DC fields detected by only $X$ phase of the RESOLUTE sequence compared to the full RESOLUTE signal where DC fields are extracted.

In the case where the outcomes are subtracted the phase would result in zero if the DC magnetic field has not changed during the correlation time and only phase due to noise would remain which would be the cause of the decoherence of the averaged signal (Eq.\,\ref{eq:Ramsey_corr_noise}). However, if a pulse addressing a target electron spin is  applied during the correlation time, the sign of the dipolar interaction would flip and the outcome of the subtracted signal would result in isolated phase of the dipolar coupling.

\begin{equation}\begin{split}
\braket{0|\psi_{\mathrm{NV}_{\hat{x}\hat{x}}}}-\braket{1|\psi_{NV_{\hat{y}\hat{y}}}}&=S(\phi=x)-S(\phi=y)\\&=\cos\left(\phi_1-\phi_2\right)=\cos\left(\omega_{dd}\right)\end{split}
\label{eq:Subtract_signal}
\end{equation}

\subsection{Low AC magnetic field sensing with RESOLUTE and noise correlation} \label{ap:RCPS-noise}

We have shown in the main text that the Ramsey correlation sequence (RESOLUTE) exhibits a coherence time longer than $T_2^*$, the coherence time of a regular Ramsey sequence. In addition, we observed an increase in coherence at specific correlation times, which we attribute to the correlation with AC magnetic fields, such as the carbon nuclear spin in the lattice. To understand the mechanism of noise correlation, we derive the expected signal of a single iteration of uncorrelated noise and correlated noise at different correlation times (correlated or uncorrelated to the noise). To account for the averaged nature of the NV sensor, we then integrate the signal with the remaining noise over a Gaussian distribution of the noise to receive the expected decaying signal.

We consider three noise sources: First, an AC field correlated noise. For this derivation, we will use the carbon nuclear spin, with $T_{\text{corr}}=nT_{\mathrm{C}}$. Second, a DC field noise fluctuating with the measurement timescale $\delta B_{\text{corr}}$. Third, an uncorrelated noise $\delta b$. The second noise, the fluctuating DC field $\delta B_{\text{corr}}$, is the main source for coherence gain after $T_2^*$. This noise has no specific time characteristics. However, if we assume that the field $\delta B_{\text{corr}}$ is constant throughout a single iteration, we can write the expected signal for both measurements ($\phi=x/y$) in Eq.\,\ref{eq:Ramsey_Corr_DC_noise_phase}.

\begin{equation}
\begin{split}
     &S(\phi=x)=\cos(\Phi_1-\Phi_2)+\cos(\Phi_1+\Phi_2)=\\
     & \\
     &\cos((\omega_{dd}+\gamma\delta b_1)\tau)+\cos((\gamma\Delta B + A_n + \gamma \delta B_{\text{corr}1})\tau)  \\
     & \\
     & S(\phi=y)=-\sin(\Phi_1-\Phi_2)+\sin(\Phi_1+\Phi_2)=\\
     & \\
     &-\cos((\omega_{dd}+\gamma\delta b_2)\tau)+\cos((\gamma\Delta B + A_n + \gamma \delta B_{\text{corr}2})\tau)
\end{split}
\label{eq:Ramsey_Corr_DC_noise_phase}
\end{equation}

The noise correlated with the measurement times' cycle is canceled when the phases of the two interaction periods are subtracted. However, it remains when they are added together with any other DC magnetic field.  In contrast to the other DC fields, this noise is not the same between iterations and even between both measurements of $\phi=x/y$. Therefore, when adding the measurements of different phases $\phi=x/y$ and adding all repetitions, we can treat the noise $\delta B_{\text{corr}}$ as a Gaussian distributed noise and integrate over the noise to result with an exponential decay with a decay constant of $\Gamma_{B_{\text{corr}}}$ (similar to Eq.\,\ref{eq:Ramsey_corr_DC_noise} with $\alpha_1=\sfrac{1}{\Gamma_{B_{\mathrm{DC}}}}$). 

The subtracted phases still have noise from any uncorrelated magnetic fluctuations $\delta b$. Therefore, the subtracted phase component after subtracting the measurements of different phases $\phi=x/y$ and averaging over all repetitions will end up with an exponential decay. However, this signal would be integrated over a different Gaussian distribution of the noise where the correlated noise is factored out and thus, the expected decay constant will be higher $\Gamma_b$ (similar to Eq.\,\ref{eq:Ramsey_corr_noise} with $\alpha_2=\sfrac{1}{\Gamma_b}$).

Finally, the AC field correlated noise has a similar mechanism to the $\delta B_{\text{corr}}$ noise, with one exception: The coherence gain has time reoccurrence for any integer multiplication of the correlation time $T_{\text{corr}}=nT_{\mathrm{C}}$\cite{reinhard2012tuning}. To derive the expected phase of a single iteration, we consider the phase accumulated due to the AC-correlated noise at the first and second interaction periods. For now, we will ignore other phase sources for simplicity. The phase in the first period will be $\Phi_1 = 2\pi (f_{\mathrm{C}}\frac{\tau}{2})$ and the phase of the second period will be $\Phi_2 = 2\pi (f_{\mathrm{C}}\frac{\tau}{2}+f_{\mathrm{C}}T_{\text{corr}})$. Together, we can write the outcome of a single iteration as the addition and subtracted phases (Eq.\,\ref{eq:Ramey_corr_AC_noise}).

\begin{subequations}
\begin{equation}
\begin{split}
     &S(\phi=x,\tau=\tau_{\mathrm{fix},T_{\text{corr}}})=\\
     & \\
     & \cos(-2\pi f_{\mathrm{C}}T_{\text{corr}}) + \cos(2\pi (\sum f_{\mathrm{C}}^1\frac{\tau}{2}+f_{\mathrm{C}}T_{\text{corr}}))  \\
     & \\
     & S(\phi=y,\tau=\tau_{\mathrm{fix},T_{\text{corr}}})=\\
     & \\
     &-\cos(-2\pi f_{\mathrm{C}}T_{\text{corr}}) + \cos(2\pi (\sum f_{\mathrm{C}}^2\frac{\tau}{2}++f_{\mathrm{C}}T_{\text{corr}}))
\end{split}
\label{eq:Ramey_corr_AC_noise}
\end{equation}
\begin{equation}
\begin{split}
    S(\phi=x/y, T_{\text{corr}})=&\pm\cos(-2\pi f_{\mathrm{C}}T_{\text{corr}}) \\
    & \begin{split}+\int &\left[\cos(2\pi(\sum f_{\mathrm{C}}\frac{\tau}{2} +f_{\mathrm{C}}T_{\text{corr}}))\right]\\ & \cdot e^{-\Gamma_{\sum f_{\mathrm{C}}}(\sum f_{\mathrm{C}})^2}d \sum f_{\mathrm{C}}\end{split} \\
    &\begin{split}= &\pm\cos(f_{\mathrm{C}}T_{\text{corr}}) \\ & + e^{\frac{\tau^2}{4\Gamma_{\sum f_{\mathrm{C}}}}}\cos(f_{\mathrm{C}}T_{\text{corr}})\end{split}
\end{split}
\label{eq:Ramey_corr_AC_average}
\end{equation}    
\begin{equation}
    S(T_{\text{corr}}) = S(\phi=x)-S(\phi=y)=\cos(f_{\mathrm{C}}T_{\text{corr}})
    \label{eq:Ramey_corr_AC_sub}
\end{equation}
\end{subequations}

Both terms have the phase of the AC signal from the correlation time, resulting in an oscillating function in $T_{\text{corr}}$ with a frequency of $f_{\mathrm{C}}$. However, the second term also has a residual phase $\sum f_{\mathrm{C}}$, which is changing between iterations. The final signal of the measurement is obtained after subtracting or adding the separate phase iterations and averaging over all repetitions. The order of these operations can be switched. Starting with averaging over all repetitions for each phase measurement $\phi=x/y$, we are adding up all coherent signals and averaging the noise signal by integrating over the signal with a Gaussian distribution of the noise (Eq.\,\ref{eq:Ramey_corr_AC_average}). 

The result is a coherently oscillating function and an exponentially-decaying oscillating function. Now, subtracting the $\phi=x/y$ measurement will result only in the oscillating function (Eq.\,\ref{eq:Ramey_corr_AC_sub}). The result in Eq.\,\ref{eq:Ramey_corr_AC_sub} is incomplete, as other noise sources, such as relaxation and fast fluctuating magnetic noise, are still causing a decay of the averaged signal and are not included in this derivation.

\section{Using adiabatic pulses to improve electron spin sensing}\label{AppendixE}

In NV magnetometry, adiabatic pulses, also called chirped pulses, have been used for improving readout in single-NV ODMR \cite{Niemeyer2013, Genov2020, Scheuer2017} or for increasing the NV ensemble response in nanodiamonds, where the orientation is not known \textit{a-priori} \cite{Rendler2017}. These are pulses where the frequency of the applied MW field is changing during the pulse with a change rate of $d\alpha/dt$. Due to the off-resonance frequency of the field, the applied pulse has an effective magnetic field $B_\text{eff}$ which is tilted from the $x$ axis of the rotating frame of the sensor \cite{Genov2020}. To apply a highly adiabatic pulse the rate of frequency change would have to be much smaller than the effective applied magnetic field (Eq.\,\ref{eq:Q factor}). The minimal adiabaticity factor can be found when the driving frequency approaches resonance, the effective field is minimal, and the rate of frequency change is maximal. At this point, it is also possible to relate the adiabaticity factor to the system parameters of Rabi frequency ($\nu$), frequency span of the sweep ($\Delta F$), and the pulse time ($t_p$), which results in the term written in Eq.\,\ref{eq:Q min},
\begin{subequations}
\begin{equation}
Q = \frac{\omega_{\mathrm{eff}}(t)}{d\alpha/dt}
\label{eq:Q factor}
\end{equation}
\begin{equation}
Q_\text{min} = \frac{2\pi \nu^2 t_p}{\Delta F} \gg 1
\label{eq:Q min}
\end{equation}
\end{subequations}
To show the advantage that adiabatic pulses have over rectangular $\pi$-pulses in electron spin detection with NV centers, we scanned shallow NV centers using a double-electron-electron-resonance (DEER) sequence~\cite{Grotz2011} to detect electron spins. We compared the signal between a $\pi$-pulse on the target electron spins and an adiabatic pulse. Each of the NV centers was scanned four times. First, we applied a DEER-frequency and a DEER-$\tau$ (see Appendixes \ref{ap:DEER}, \ref{ap:DEER_t}) with a $\pi$-pulse addressing the electron spins, where the pulse duration was estimated from the NV center's Rabi frequency. Second, we applied again the DEER-frequency and DEER-$\tau$ changing the $\pi$-pulse to a chirped pulse addressing the electron spins. To make the signals comparable, we used a chirped pulse with a quality factor $Q=5$ and duration of  1.6\,$\upmu$s that resulted in a chirp span of 2.5\,MHz where the spectral width of the $\pi$-pulse was 2.2\,MHz. We examined four NV centers in the diamond at an average depth of 10\,nm. All examined NVs had a $T_2$ longer than 10\,$\upmu$s to ensure sufficient sensing volume. 

\begin{figure}[t]
\includegraphics[width=0.9\columnwidth]{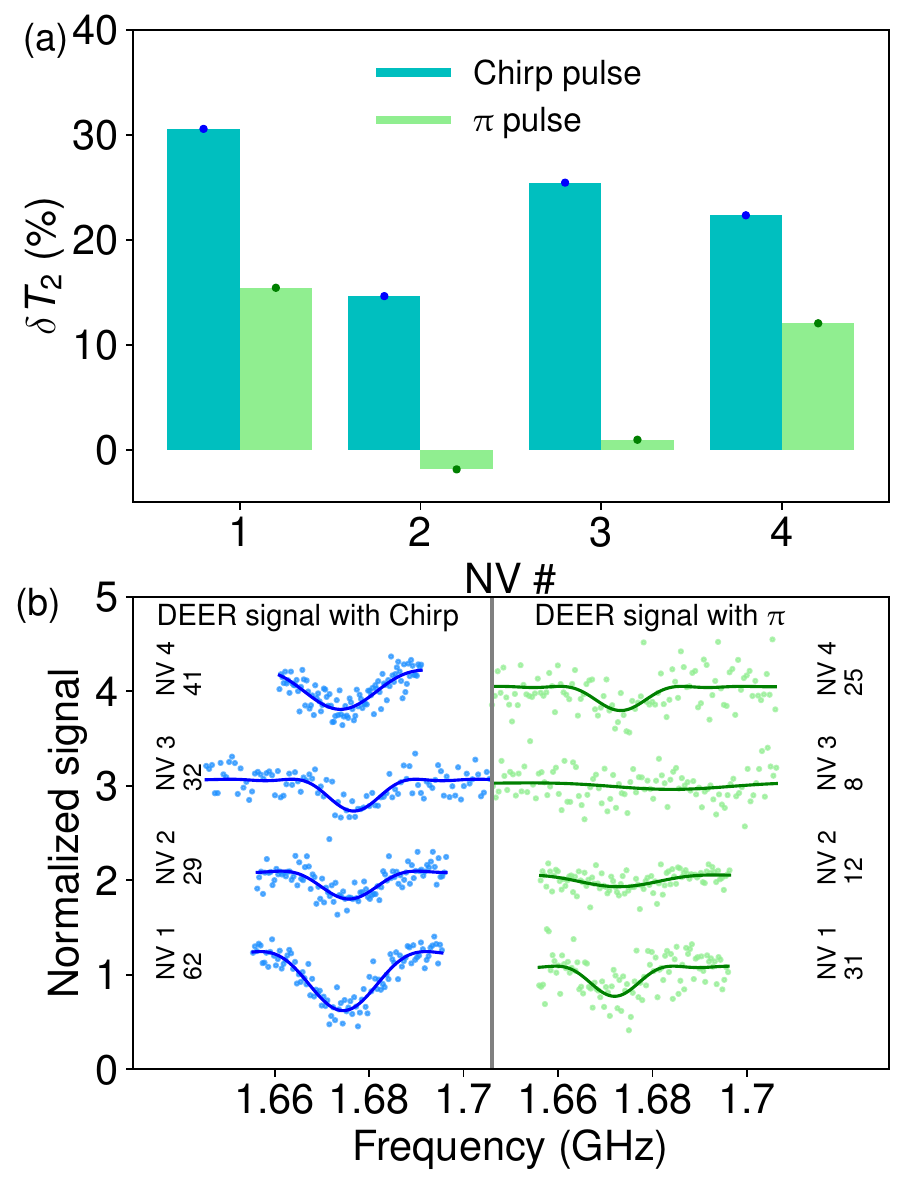}
\caption{\textbf{Chirped vs.\, rectangular $\pi$-pulses for DEER in RESOLUTE}. (a) Comparing the DEER contrast of DEER when using chirped pulses (left, blue) to the more conventional $\pi$-pulses (right, green) for NVs 1-4. There is a clear improvement in contrast when using chirped pulses, which sometimes, e.g. NV3, results in detecting the presence of the target electron spin using a chirped pulse compared to observing no signal at all with the rectangular $\pi$-pulse. (b) A quantitative calculation of the relative change in coherence time with and without a pulse using chirped pulses (blue) and $\pi$-pulses (green), with $\delta T_2=T_{2,\text{w/}} - T_{2,\text{w/o}}$. Here, too, the chirped pulses offer a larger contrast.}
\label{fig:Chirp}
\end{figure}

Results of the DEER-frequency scans are plotted in Fig.\,\ref{fig:Chirp}. All scanned data were fitted to a single Gaussian function to estimate the contrast of the DEER signal. The normalized contrast (\%) of each curve is specified under the NV label and fit plotted in Fig.\,\ref{fig:Chirp}. The left column (blue curves) are scans taken with a chirped pulse on the electron spin, while the right column (green curves) are taken with a $\pi$-pulse. It is clear from the fit that the contrast with a chirped pulse was higher for all scanned NVs. The fitted width of the Gaussian was, on average, $15\,\mathrm{MHz}$, indicating that the broadening is not solely rising from the pulse span and is still narrower than the linewidth reported for similar experiments~\cite{Dwyer2022, Mamin2012}.

We scanned all four NV centers with a DEER-$\tau$ sequence. To make the comparison between a chirp and $\pi$-pulses, we took each scan twice: Once with a pulse on the electron spin resonance frequency and then also with a pulse that is detuned 15 MHz from the electron spin resonance frequency, making it completely off-resonance from the detected signal at the DEER frequency. The $\pi$-pulse was with a length of $440\,\mathrm{ns}$, resulting in a width of $\sim2.2\,\mathrm{MHz}$. The chirped pulses were all with $Q=5$ and length of $2\,\upmu\mathrm{s}$, resulting in a span of $\sim3\,\mathrm{MHz}$. All scans were fitted to a stretched decay exponent of the form $S(\tau)=Ae^{-(\sfrac{\tau}{\Gamma})^\beta}+c$. A pronounced signal of the electron spins was detected with the chirp pulse while a faint or no signal at all was detected with the hard $\pi$-pulse, as expected.

\section{Fisher information analysis of RESOLUTE}\label{AppendixTheory}

Let us begin by comparing the filter functions from a Ramsey sequence, a Hahn Echo, and a RESOLUTE measurement, to then proceed to understand frequency selectivity with RESOLUTE through an exact Fisher information analysis. Assuming throughout that we are trying to detect a pure tone signal $A\sin(\omega t + \varphi)$, the phase that the qubit accumulates is the integral of the signal during the sensing time of the sequence. Then, for a Ramsey measurement of duration $\tau_R$ we have that
\be\label{eqSI:accumRamsey}
\Phi_{\text{R}}(\tau_R,\omega) = A\tau_R\cos\left(\frac{\omega \tau_{\text{R}}}{2} + \varphi\right)\sinc\left(\frac{\omega \tau_{\text{R}}}{2}\right),
\ee
which peaks at zero frequency owing to the $\sinc$ term, and allows a small window of frequency detection by using the displacement that the signal's phase $\varphi$ allows. Averaging over many uncorrelated measurements with different phases destroys the frequency sensitivity and explains why the Ramsey sequence is most suited for DC sensing. 

Improving on the Ramsey sequence, a Hahn Echo adds a $\pi$ rotation on the Bloch sphere in the middle of the sensing period to eliminate low frequency noise and center the frequency sensitivity peak around the inverse sequence characteristic time $\tau_{\text{H-E}}$, as reflected in the accumulated phase, which reads  
\begin{multline}\label{eqSI:accumHE}
\Phi_{\text{H-E}}(\tau_{\text{H-E}},\omega) = A\tau\sin\left(\frac{\omega \tau_{\text{H-E}}}{2} + \varphi \right) \times \\  \sin\left(\frac{\omega \tau_{\text{H-E}}}{2}\right)\sinc\left(\frac{\omega \tau_{\text{H-E}}}{4}\right),
\end{multline}
where matching the inter-pulse delay time $\tau_{\text{H-E}}$ to the signal's frequency, phase accumulation can be maximized for the specific target signal, providing with an AC spectrometer.

Similar to a Hahn Echo, and following Eq.\,\ref{eq:Subtract_signal}, the global accumulated phase on a RESOLUTE measurement is 
\begin{multline}\label{eqSI:accumphase}
\Phi(\tau,\tcorr,\omega) = A\tau\sin\left(\frac{\omega T_{\text{corr}}}{2} + \frac{\omega \tau}{2} + \varphi \right)\times \\ \cos\left(\frac{\omega T_{\text{corr}}}{2} + \frac{\omega \tau}{4}\right)\sinc\left(\frac{\omega \tau}{4}\right).
\end{multline}
Comparing Eq.\,\ref{eqSI:accumphase} for RESOLUTE with Eq.\,\ref{eqSI:accumHE} provides intuition about the frequency detection condition for RESOLUTE, in which the role of frequency matching that in Hahn Echo is played by the inter-pulse delay time $\tau_{\text{H-E}}$ corresponds in RESOLUTE to $T_{\text{corr}} + \tau$, endowing the RESOLUTE sequence with a larger scope, as there are two controllable time variables that can be used to achieve resonance with the target signal.  

Further, if we compare Eq.\,\ref{eqSI:accumphase} with the phase that is accumulated on a Ramsey sequence in Eq.\,\ref{eqSI:accumRamsey} we observe that it is sensitivity to the target signal's phase $\varphi$ that confers the RESOLUTE sequence with the ability to detect oscillating signals. A typical Ramsey experiment comprises many uncorrelated measurements which are described by a phase-average of Eq.\,\ref{eqSI:accumRamsey}, thus destroying the frequency selectivity \cite{Maze2012}. In RESOLUTE, even though one relies on Ramsey sensing periods, these are pairwise separated by a \textit{fixed} delay time $\tcorr$, meaning that, upon averaging, the signal's phase ($\varphi$) correlations survive, thereby enabling the sensitivity to oscillating signals. A similar principle is exploited by phase-sensitive sequences such as those proposed in Refs.\,\onlinecite{Herbschleb2022, Oviedo2024}, where phase correlations survive by having Ramsey signal acquisition periods evenly separated in time, and correlating them during post-processing. The key difference that RESOLUTE brings about is that, by introducing an extra time-variable, namely $T_{\text{corr}}$, RESOLUTE provides greater flexibility to match the target's signal frequency, meaning that rather than relying on continuous measurement accumulation, RESOLUTE can be tailored to specific signals. The resemblance to a Hahn Echo explains the frequency selectivity in RESOLUTE, which in turn rationalizes the increased coherence time observed in Fig.\,\ref{fig:T2_compare}a, a feature not present in phase-sensitive approaches that track the evolution of any signal whose period is longer than the sensing time.

Further understanding on the reason why a RESOLUTE sequence allows AC sensing despite using Ramsey phase accumulation periods comes from a filter function analysis, which corresponds to the Fourier transform of the qubit response function to the sequence of pulses or, alternatively, the phase-averaged square of the accumulated phase, $\langle \Phi^2 \rangle_\varphi$. The filter function reflects the signal accumulation on the probe, and directly relates theory with experimental results.
For Ramsey, such phase averaging yields, as expected,

Additional insight on the frequency selectivity of RESOLUTE can be obtained by resorting to the concept of filter functions, which corresponds to the Fourier transform of the qubit response function to the sequence of pulses or, alternatively, the phase-averaged square of the accumulated phase, $\langle \Phi^2 \rangle_\varphi$. The filter function reflects the signal accumulation on the probe, and directly relates theory with experimental results. For Ramsey, the filter function is 
\be
\langle \Phi_R(\tau_R,\omega)^2 \rangle_\varphi = \frac{A^2\tau_R^2}{2} \sinc\left(\frac{\omega\tau_R}{2} \right),
\ee
which, as expected, peaks at zero frequency. For a Hahn Echo, frequency sensitivity survives the phase averaging, yielding a filter function that reads
\be
\langle \Phi_{\text{H-E}}(\tau_{\text{H-E}},\omega)^2 \rangle_\varphi = \frac{A^2\tau_{\text{H-E}}^2}{2}\sin^2\left(\frac{\omega \tau_{\text{H-E}}}{2}\right)\sinc^2\left(\frac{\omega \tau_{\text{H-E}}}{4}\right).\label{eqSI:HEff}
\ee

In the case of RESOLUTE, the filter function reads
\begin{equation}
\label{eqSI:ff}
\begin{split}
\langle \Phi(\tau,\tcorr,\omega) \rangle_\varphi = \frac{A^2\tau^2}{2}\cos^2\left(\frac{\omega T_{\text{corr}}}{2} + \frac{\omega \tau}{4}\right)\sinc^2\left(\frac{\omega \tau}{4}\right),
\end{split}
\end{equation}
which imitates the Hahn Echo in Eq.\,\ref{eqSI:HEff} and shows that frequency sensitivity is maintained despite using Ramsey phase acquisition periods. We validate the theoretical filter function for RESOLUTE by explicitly calculating it simulating detection of three frequencies $\omega_2 = 2\pi\times 117$\,kHz, $\omega_1 = 2\pi\times 71$\,kHz and $\omega_2 = 2\pi\times 53$\,kHz using the same parameters as in the experiment shown in Fig.\,\ref{fig:T2_compare}b on the main text, demonstrating that the spectral peaks are correctly reproduced. 

\begin{figure}
\includegraphics[width=\columnwidth]{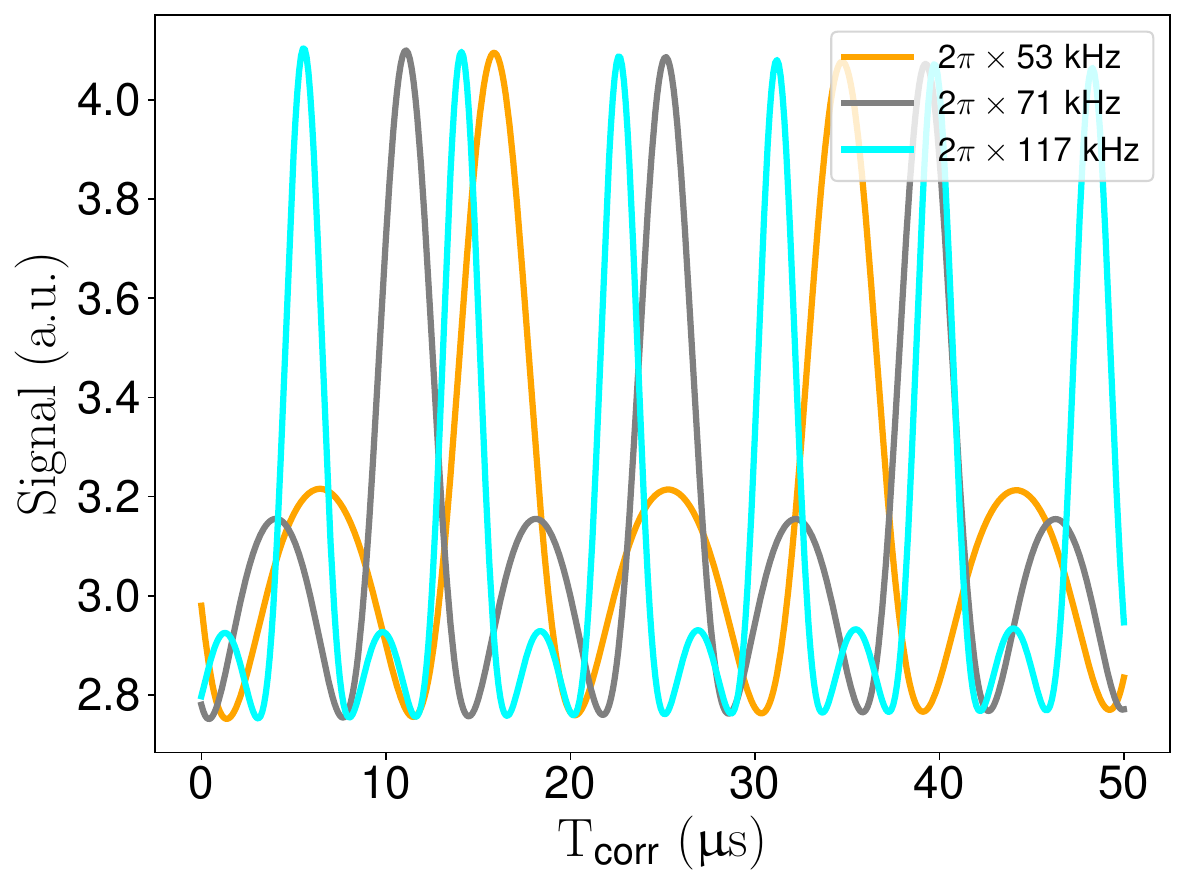}
\caption{Filter function Eq.\,\ref{eq:ff} for detection of three frequencies $\omega_1 = 2\pi\times 53$\,kHz, $\omega_2 = 2\pi\times 71$\,kHz and $\omega_2 = 2\pi\times 117$\,kHz, simulating detection with a probe featuring a $T_2^p = 5.1\,\upmu\text{s}$ and $T_1 = 1000\,\upmu\text{s}$, with interaction time $\tau = 5.5\,\upmu\text{s}$ as a function of the correlation time $T_{\text{corr}}$, showing the resonant peaks at $\omega  T_{\text{corr}.} \approx n$.}\label{figSI:theoryFF}
\end{figure}

Finally, we perform a Fisher information analysis where we compute the exact Fisher information for each block on a RESOLUTE measurement and analyze it aided by the approximate formula provided on the main text, for which we give further details on its derivation. Considering the frequency $\omega$ of a pure tone signal as the target parameter to estimate from the experiment, the Fisher information for a measurement is
\be\label{eqSI:singleFI}
\begin{split}
i_\omega &= \sum_{X\in\left\{0,1\right\}} \frac{1}{P(X|\omega)}\left[ \frac{dP(X|\omega)}{\omega}\right]^2 \\&= \frac{1}{P(0|\omega)[1-P(0|\omega)]}\left[ \frac{dP(0|\omega)}{\omega}\right]^2,
\end{split}
\ee
where we use the fact that for a binary measurement in a qubit probe $P(1|\omega) = 1-P(0|\omega)$. We can write this probability in terms of the phase in Eq.\,\ref{eqSI:accumphase} that the qubit accumulates during one measurement block of RESOLUTE. The key question is to realize that, to exactly calculate the Fisher information we need to take into account that for each block the probability is slightly different. Thus, for the $XXX\pm X$ we get a probability 
\be
P(0|\omega) = \frac{1}{2}\left(1 \pm A\tau e^{-\frac{\tau}{T_2^p} - \frac{\tcorr}{T_1}}\cos(\phi_2-\phi_1)\right),
\ee
while for the $XYY\pm X$ we obtain that
\be
P(0|\omega) = \frac{1}{2}\left(1 \pm A\tau e^{-\frac{\tau}{T_2^p} - \frac{\tcorr}{T_1}}\cos(\phi_1)\cos(\phi_2)\right),
\ee
with $\phi_{1,2}$ the phases accumulated during the first (second) sensing periods. Then, each block yields a slightly different Fisher information.

All figures below are drawn with data calculated numerically using the exact Fisher information for RESOLUTE, that is, considering each block as a distinct entity with its own Fisher information. To derive Eq.\,\ref{eq:approxFI} on the main text, we use the following approximations. First, note that owing to the symmetry and the square in Eq.\,\ref{eqSI:singleFI}, the sign change in the different probabilities which is a result of flipping the last ($X$) $\frac{\pi}{2}$ pulse becomes irrelevant, and the Fisher information becomes identical for the $XXX\pm X$ blocks and the $XYY\pm X$ blocks. The former yield
\be\begin{split}\label{eqSI:FI1}
&i_\omega(XXX\pm X) \approx \\& \frac{2\sin^2\Phi^-}{\left[\exp\left(2\tau/T_2^p + 2T_{\text{corr}}/T_1 \right)-\cos^2\Phi^-\right]}\left(\frac{d\Phi^-}{d\omega}\right)^2,
\end{split}\ee
with $\Phi^- = \phi_2 - \phi_1$. On the other hand, the latter read
\be\begin{split}\label{eqSI:FI2}
&i_\omega(XYY\pm X) \approx \\&\frac{2}{\left[\exp\left(2\tau/T_2^p + 2T_{\text{corr}}/T_1 \right)-\cos(\phi_1)\cos(\phi_2)\right]} \\& \left(\sin(\phi_1)\cos(\phi_2)\frac{d\phi_1}{d\omega} + \cos(\phi_1)\sin(\phi_2)\frac{d\phi_2}{d\omega}\right)^2.
\end{split}\ee

In the small phase accumulation limit $\phi_1,\phi_2 \ll 1$, relevant for typical experiments where $\omega \tau \ll 1$, $\sin(\phi_i) \approx \phi_i$ and $\cos(\phi_i) \approx 1$. Moreover, in this limit $\phi_1 \approx \phi_2$ and then $\frac{d(\phi_1 + \phi_2)}{d\omega} \ll \frac{d(\phi_2 - \phi_1)}{d\omega}$. All together it means that the Fisher information in the $XYY\pm X$ blocks is in this limit similar to that of the $XXX\pm X$ blocks, and then the total Fisher information on a RESOLUTE measurement can be well approximated by four times that of an $XXXX$ block. 

The small phase accumulation limit also corresponds to having a small $\sinc$ term in Eq.\,\ref{eqSI:accumphase} for narrower spectral lines, and means that $\sinc \approx 1$ for good signal accumulation. Then, we can approximate $\sinc' \approx 0$ and remove the terms depending on this derivative from the Fisher information approximation. This means that the derivative with respect to the frequency of the accumulated phase approximates as
\be\label{eqSI:phasederiv}
\left(\frac{d\Phi}{d\omega}\right)^2 \approx \frac{A^2\tau^2\tilde{T}^2}{4}\cos^2\left( \omega\tilde{T}+\frac{\omega \tau}{4}+\varphi\right)\sinc^2\left(\frac{\omega \tau}{4}\right),
\ee
where $\tilde{T} = T_{\text{corr}} + \tau/2$.

Finally, the first factor in Eq.\,\ref{eq:FI1}, namely 
\be
\frac{\sin^2\Phi}{\left[\exp\left(2\tau/T_2^p + T_{\text{corr}}/T_1 \right)-\cos^2\Phi\right]},
\ee
can be approximated by its upper bound $\exp\left(-2\tau/T_2^p - T_{\text{corr}}/T_1 \right)$, which reflects the limits on the different measurement times. Then, we get the Fisher information approximate formula 
\be
\begin{split}\label{eqSI:approxFI}
i^{\mathrm{approx}}_\omega &= 8A^2\tau^2\tilde{T}^2\exp\left(-2\tau/T_2^p - T_{\text{corr}}/T_1 \right) \\& \times\cos^2\left( \omega\tilde{T}+\frac{\omega \tau}{4}+\varphi\right)\sinc^2\left(\frac{\omega \tau}{4}\right).
\end{split}
\ee

\begin{figure}
\includegraphics[width=\columnwidth]{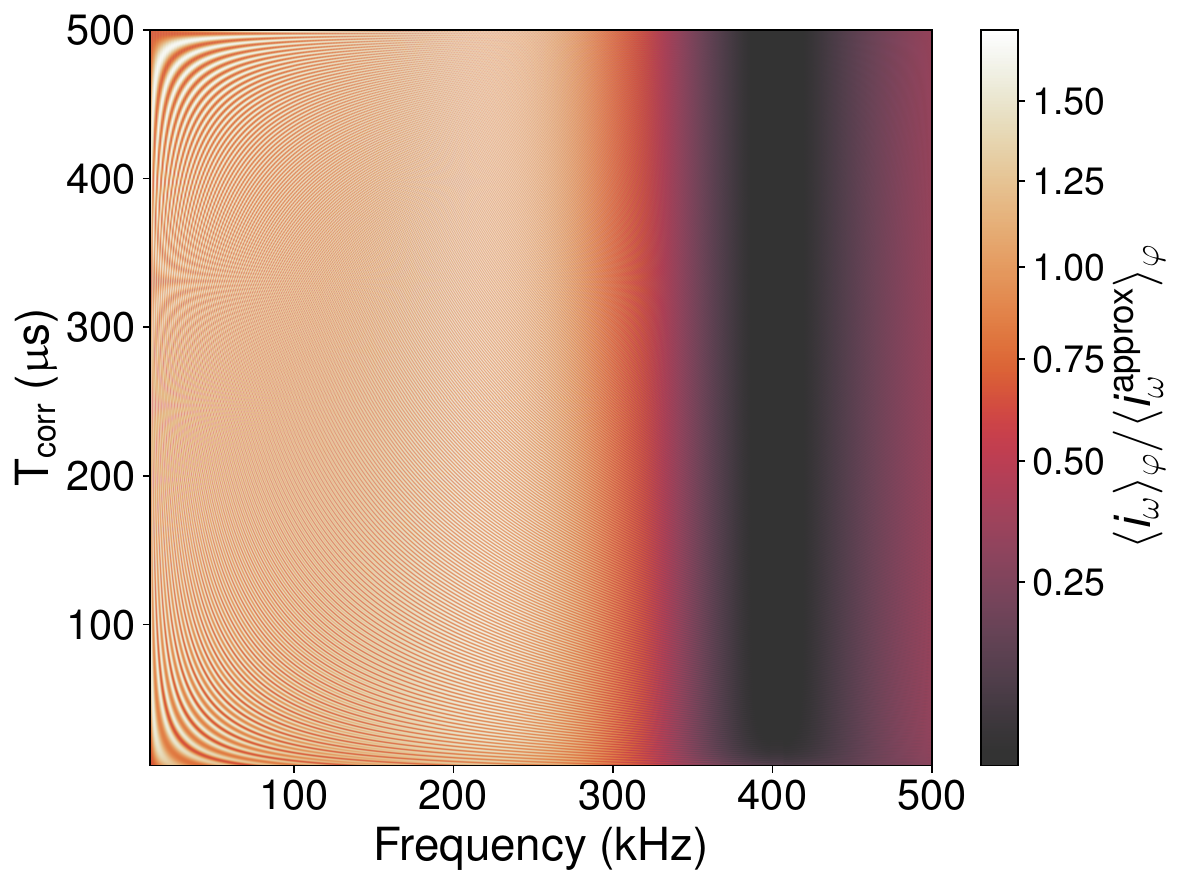}
\caption{Fisher information ratio between the exact expression for an RESOLUTE measurement and the approximate formula in Eq.\,\ref{eq:approxFI}, calculated at $\tau = 5 \,\upmu\text{s}$, for a wide range of frequencies and correlation times, considering a $T_2^p = 5.1 \,\upmu\text{s}$ and a $T_1 = 1000 \,\upmu\text{s}$. The approximate formula underestimates RESOLUTE capabilities at low frequencies, representing thus a lower bound for an experiment, while for the chosen $\tau$ it overestimates the Fisher information at high frequencies, for which anyways a shorter $\tau$ should be preferred, in whose case the overestimation would is displaced, as it stems from the different Ramsey fringe description between the exact and approximate formulas.}\label{figSI:theory:exactVsApprox_tau5}
\end{figure}

A numerical comparison of Eq.\,\ref{eq:approxFI} to the exact Fisher information ---calculated as the sum of the Fisher information for each of the measurement blocks in a given RESOLUTE sequence-- can be seen in Fig.\,\ref{figSI:theory:exactVsApprox_tau5}, where for a sensing time $\tau = 5\,\upmu\text{s}$, we observe that the approximate formula reproduces fairly well the exact results for most of the relevant target spectrum, only slightly underestimating the capabilities of RESOLUTE at very low frequencies, where the exact result is further from the upper bound of the first factor in Eq.\,\ref{eq:FI1}, while overestimating them at relatively high frequencies, where the $\phi_1 \approx \phi_2$ approximation breaks down. 

\begin{figure}
\includegraphics[width=\columnwidth]{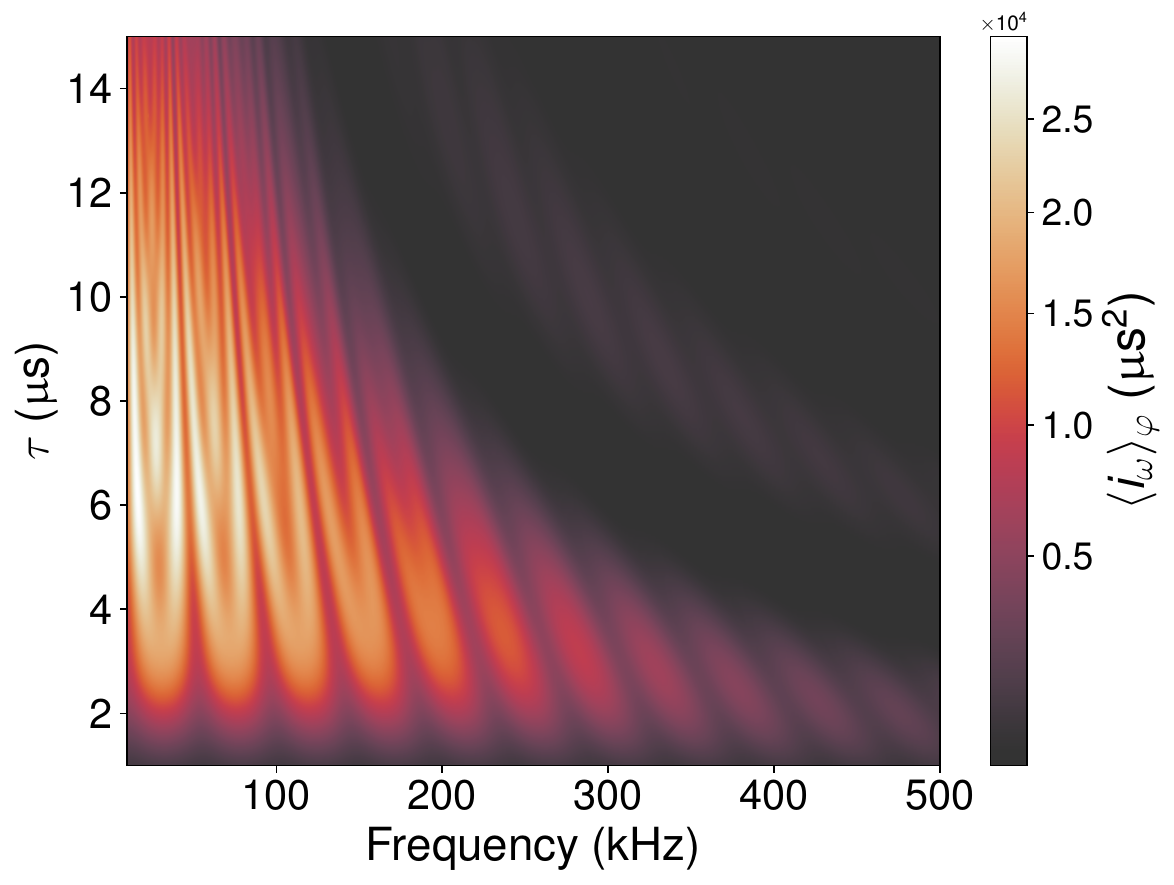}
\caption{Exact $i_\omega$ for RESOLUTE as a function of the frequency and the sensing time $\tau$, for a correlation time $T_{\text{corr}} = 100\,\upmu\text{s}$, and with decoherence and relaxation times $T_2^p = 5.1\,\upmu\text{s} $ and $T_1 = 1000\,\upmu\text{s}$, respectively. At large sensing time, the Fisher information decreases dramatically due to the exponential suppression from decoherence noise, while a large $\omega\tau$ suppresses the Fisher information due to the $\sinc$ term. A much suppressed secondary sensing fringe can be appreciated. Note that we consider the contrast and the target signal amplitude both equal to one. In a real experiment they would contribute a constant factor much smaller than one, further diminishing the Fisher information in each measurement, but that can be compensated for with more measurements. }\label{figSI:theory:omega_tau_FI_corr}
\end{figure}

Analyzing RESOLUTE for the  best experimental setup for a given target signal, in Fig.\,\ref{figSI:theory:omega_tau_FI_corr} we choose a correlation time $\tcorr = 100 \,\upmu\text{s}$ and explore, with a numerically exact calculation, the Fisher information for an RESOLUTE sequence as a function of the frequency and the sensing time. Having chosen a larger $\tcorr \gg \tau$ we observe the clearly defined fringes that the resonance condition produces, in which $\tau$ causes a small modulation which becomes more pronounced and complex for larger sensing time, with sidebands appearing and which bespeak of the flexibility that RESOLUTE protocol has for targeting and isolating spectral components. The most important observation is the confirmation that low frequencies favor $\tau \sim T_2^p$, chosen here at $T_2^p = 5.1 \,\upmu\text{s}$, while larger frequencies require from shorter phase acquisition times and, consequently, due to the $\tau^2$ factor in Eq.\,\ref{eq:approxFI}, are proportionally more difficult to detect, which is a direct consequence of the fact that we are using a Ramsey protocol, reflected in the $\sinc$ term on Eq.\,\ref{eqSI:approxFI}.

\begin{figure}
\includegraphics[width=\columnwidth]{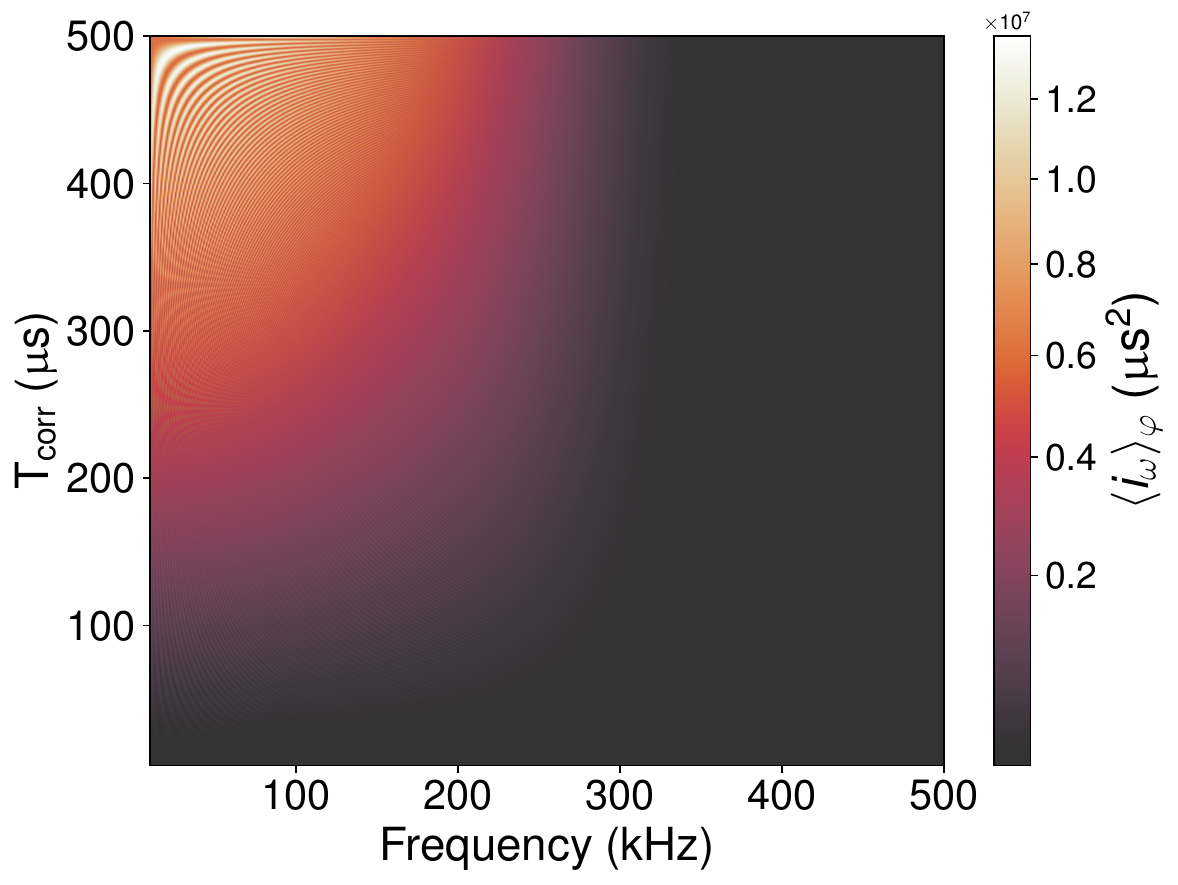}
\caption{Fisher information $i_\omega$ for a sensing time $\tau = 5\,\upmu\text{s} $, and with $T_2^p = 5.1 \,\upmu\text{s}$ and $T_1 = 1000\,\upmu\text{s}$. A large $\tcorr < T_1$ provides larger information about the target frequency, facilitating estimation, particularly at low frequencies for which a large $\tcorr$ is needed to meet the resonance condition.  Note that we consider the contrast and the target signal amplitude both equal to one. In a real experiment they would contribute a constant factor much smaller than one, further diminishing the Fisher information in each measurement, but that can be compensated for with more measurements.. }\label{figSI:theory:omegaTcorrFI_tau_5}
\end{figure}

We also study the ability that is provided to RESOLUTE by modulating $\tcorr$. In Fig.\,\ref{figSI:theory:omegaTcorrFI_tau_5}, for a phase acquisition time $\tau = 5$ which favors low frequencies, we explore the Fisher information as a function of the frequency $\omega$ and the correlation time. In this case, the main observation is that the larger the correlation time, the better a lower frequency will be detected. The reason is that stretching the correlation time to its maximum permits witnessing more oscillations of the target signal, which makes frequency estimation easier through, e.g. least squares fitting. Note that here we are assuming that $\tcorr$ is just limited by $T_1$, when in a realistic experimental scenario, harder limits such as target signal inherent decay or experimental constraints might exist which impose a tighter limit on $\tcorr$. Nevertheless, as the $\tilde{T}^2$ factor in Eq.~\ref{eqSI:approxFI} demonstrates, the optimal strategy is to always use the largest $\tcorr$ possible.

\begin{figure}
\includegraphics[width=\columnwidth]{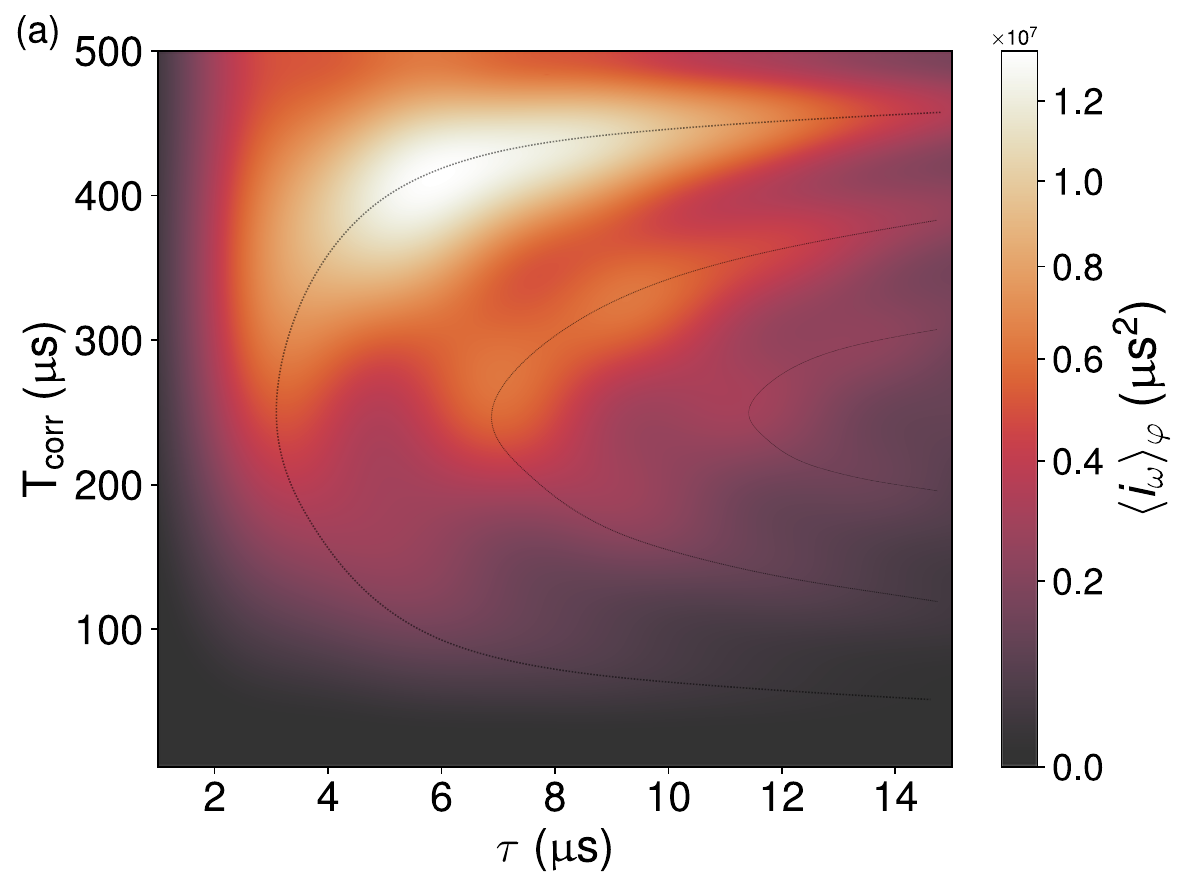}
\includegraphics[width=\columnwidth]{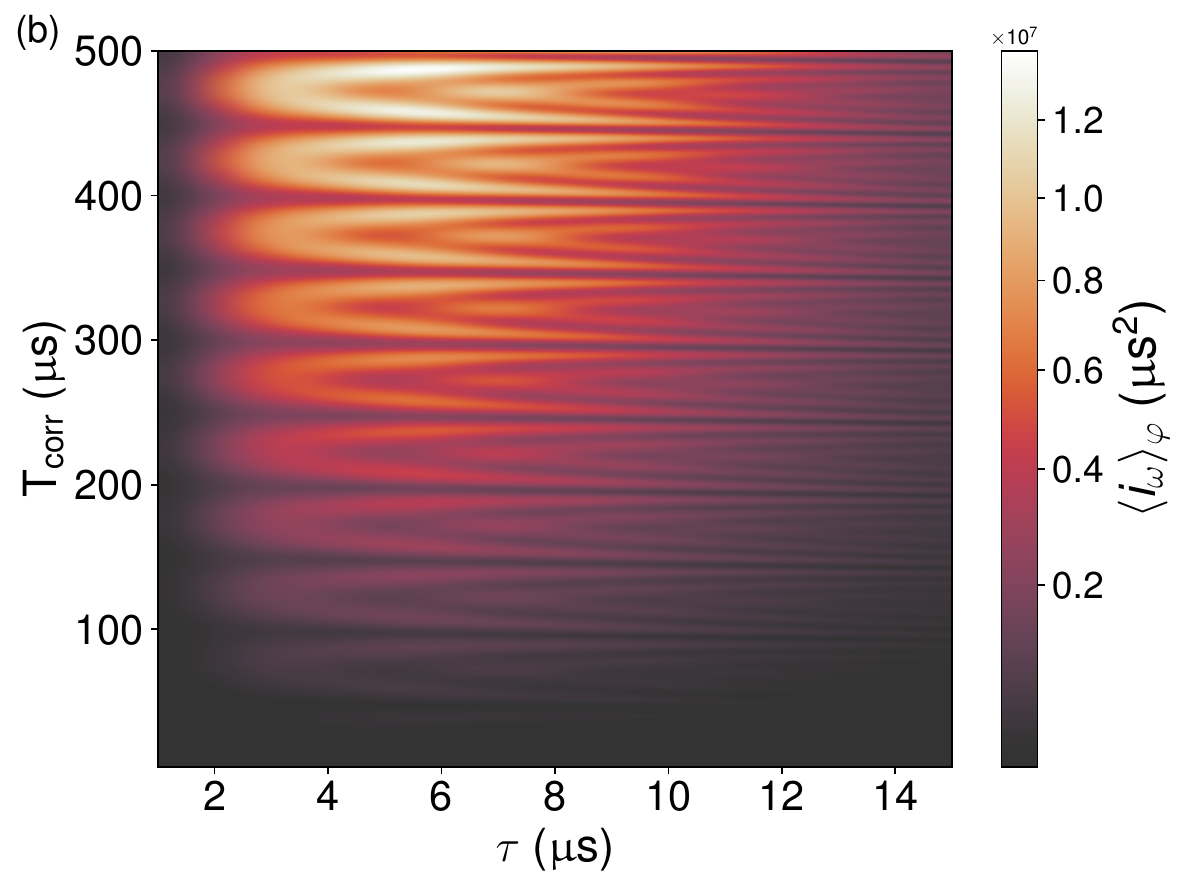}
\caption{Fisher information $i_\omega$ as a function of the sensing time $\tau$ and correlation time $\tcorr$ for two different frequencies $\omega$ = 10 (2$\pi\times$ kHz) in (a) and $\omega$ = 100 (2$\pi\times$ kHz) in (b). In both cases, a larger correlation time produces better estimation, while the optimal sensing time matches the decoherence chosen for the calculation, namely $T_2^p = 5.1 \,\upmu\text{s}$. Note that for a low frequency, the resonance condition becomes less critical, as shown by the smeared fringes appearing as a consequence of the broader $\sinc$ term, and due to the fact that for low frequencies few oscillations appear even for a large $\tcorr$. For larger frequencies, the resonance condition must be carefully balanced to ensure optimal detection, as shown by the sharper features in (b). 
Note that this figure does not take into account the signal amplitude $A$ or the signal contrast $C$, which would decrease the Fisher information further and which here are considered to be equal to one.}\label{figSI:theory:tauTcorrFIomega}
\end{figure}

Finally, we explore the inter-dependence between $\tcorr$ and $\tau$ for detection of two frequencies of 10 (2$\pi\times$ kHz) in Fig.\,\ref{figSI:theory:tauTcorrFIomega}(a) and 100 (2$\pi\times$ kHz) in Fig.\,\ref{figSI:theory:tauTcorrFIomega}(b). In both cases we observe that the optimal $\tau$ is approximately $T_2^p$, chosen in the figure to be $T_2^p = 5.1 \,\upmu\text{s}$, showing that the coherence survival enlargement that RESOLUTE provides is crucial for its ability to target small frequencies. Moreover, in both cases, increasing the sensing time leads to the formation of two bands or branches in $\tcorr$, which is useful to note for when the correlation time permitted by the experimental conditions is limited to $\tcorr \ll T_1$, in whose case carefully balancing the sensing time $\tau$ to meet the resonance condition for $\tilde{T}$ is key, and it might be advantageous to choose a shorter $\tau$, particularly at low frequencies.

\bibliography{RC_paper}

\end{document}